\documentclass[prd,aps,10pt,a4paper,showkeys]{revtex4-2}

\usepackage[english]{babel}
\usepackage{amsmath,amssymb,amsmath}
\usepackage{dcolumn}
\usepackage{bm}
\usepackage[margin=1.5cm]{geometry}
\usepackage{subcaption}
\usepackage[pdftex]{graphicx}
\usepackage{mathtools}
\usepackage[T1]{fontenc}
\usepackage{microtype}
\usepackage[shortlabels]{enumitem}
\usepackage{tabularx}
\usepackage{multirow}
\usepackage[hidelinks,breaklinks]{hyperref}
\usepackage[capitalise]{cleveref}

\usepackage{xcolor}
\usepackage{orcidlink}

\definecolor{rose}{rgb}{1.0, 0.0, 0.5}

\hypersetup{
    colorlinks,
    linkcolor={rose},
    citecolor={green!60!black},
    urlcolor={cyan!50!blue}
}

\begin{document}

\title{Influence of Generalized Ghost Dark Energy on Wormhole Geometry}

\author{Soubhik Paramanik \orcidlink{0009-0001-7774-6228}}
\email{spsoubhik@gmail.com}

\author{Anamika Kotal \orcidlink{0009-0009-8426-3206}}
\email{kotalanamika31@gmail.com}
\author{Ujjal Debnath \orcidlink{0000-0002-2124-8908}}
\email{ujjaldebnath@gmail.com}

\affiliation{Department of Mathematics, Indian Institute of Engineering\\
Science and Technology, Shibpur, Howrah-711 103, India.}


\begin{abstract}
\noindent
    This article presents a wormhole solution constructed from a generalized ghost dark energy (GGDE) source in general relativity. At first, a brief review of GGDE is presented along with the necessary mathematical frameworks. We employed the Markov Chain Monte Carlo (MCMC) technique to constrain the free parameters of the model using the CC+BAO, $Pantheon^+$, and their combined datasets. Furthermore, the Akaike Information Criterion (AIC) and Bayesian Information Criterion (BIC) were used to statistically assess the model's performance and determine its level of acceptance. Next, the basics of wormhole geometries along with the thin-shell formalism are explained in detail. After that, three different wormhole solutions associated with three different choices of the redshift function are presented ,and their various geometric properties as well as energy conditions are studied graphically. Finally, the stability of these thin--shell structures is examined analytically as well as graphically by studying the associated effective potential.

\end{abstract}


\maketitle


\section{Introduction}\label{sec1}

That the Universe's expansion is accelerating ranks among the most unforeseen and consequential findings in observational cosmology. The first hints of this behavior came from Type Ia supernova studies \cite{Riess_1998, Perlmutter_1999}, and were subsequently corroborated by measurements of the cosmic microwave background, the large-scale distribution of galaxies, and baryon acoustic oscillations. Together, these results point to a repulsive component, generally termed dark energy, as the dominant constituent of the cosmic energy budget that drives this acceleration.

While the cosmological constant $\Lambda$ offers the simplest account of this acceleration within general relativity, it runs into serious theoretical difficulties, chief among them the fine-tuning and coincidence problems \cite{Weinberg_1989}. A variety of dynamical-field alternatives have been proposed to address these issues. One such candidate, the ghost dark energy (GDE) proposal, has attracted attention because it is rooted in quantum chromodynamics (QCD), arising from the Veneziano ghost field's contribution in a curved spacetime setting \cite{ARAKI2009343, MINKEVICH2009423}. Its generalized version, the generalized ghost dark energy (GGDE) framework, allows the energy density to depend on the Hubble parameter $H$ through both a linear term and a quadratic correction in $H^2$, yielding a more adaptable model that fares better against observations \cite{Cai_2012}.

Alongside this, general relativity (GR) also admits a class of solutions known as wormholes: hypothetical tunnels or bridges linking otherwise separate regions of spacetime. Both wormholes and black holes stand out as fascinating, unorthodox solutions of Einstein's general relativity. Whereas black holes now enjoy strong observational backing \cite{Abbott_2016a, Abott_2016b, Akiyama_2019}, wormholes have yet to be directly observed and so remain a speculative possibility. A detailed assessment of how plausible such objects are can be found in \cite{KHATSYMOVSKY1994}. Conceptually, one pictures a wormhole as a bridge-like structure made possible by a non-trivial spacetime topology within GR \cite{Visser_1995}, providing a hypothetical shortcut between two points in the cosmos and, in some regimes, mimicking black-hole properties \cite{Damour_2007, Lemos_2008}. The two connected regions could lie enormous cosmic distances apart, or even in different eras altogether, which raises the theoretical prospect of wormholes serving as a means of time travel \cite{metric}. The concept can be traced back to Flamm \cite{Flamm2015}, who first touched on it while studying the isometric embedding of the Schwarzschild metric. Some years later, seeking to remove the singularity present in the Schwarzschild geometry, Einstein and Rosen put forward a similar construction that has since become known as the Einstein--Rosen bridge \cite{Einstein_1935}. The term ``wormhole'' was coined afterward by Wheeler and Misner in 1957 \cite{MISNER_1957, Wheeler_1955}. A major turning point came in 1988, when Morris and Thorne worked out the theoretical framework for a traversable wormhole \cite{metric} -- a passage roomy enough for macroscopic objects to pass through in either direction. This progress, however, carried a serious price: it required breaking a cornerstone assumption of classical physics. More precisely, traversable wormholes cannot satisfy the null energy condition (NEC) of general relativity, meaning that sustaining their geometry and stability calls for exotic matter -- material whose behavior departs from that of ordinary matter.

Ever since wormholes were first conceived, a great deal of theoretical work has gone into understanding their geometric makeup. Given that traversable wormholes necessarily violate the NEC in GR, a significant research effort has gone into finding ways around this obstacle. Rahaman \textit{et al.}~\cite{RAHAMAN2006}, for example, showed that a wormhole geometry could be built using only a vanishingly small amount of phantom energy. Halder \textit{et al.}~\cite{Halder_2019} carried this further within the Rastall gravity paradigm, in search of configurations that honor the NEC. In a similar spirit, Banerjee \textit{et al.}~\cite{Banerjee_2021} used the $f(Q)$ gravity framework to assess whether such geometries are feasible. Chakraborty \textit{et al.}~\cite{Chakraborty_2008} additionally examined wormholes in the Braneworld scenario, drawing on several non-phantom matter fields. Still other studies have looked instead to alternative, non-standard formulations of gravity. Galileon gravity \cite{DAS_2024}, for instance, has served as one such setting for wormhole solutions. A large body of further work has explored wormhole structures across various modified gravity theories; see \cite{MUSTAFA_2022, Tayde_2023, Rosa_2022, Sharif_2014, Son_2011, Das_2023, KUMAR_2024, Malligawad_2024, CHAUDHARY_2024_46, Errehymy_2024} and references therein. Separately, how exotic fluids such as Chaplygin gas affect wormhole formation and stability has been the subject of several works, including \cite{Mukherjee_2024, Paramanik_2024, Paramanik2026Rastall, Sharif2014_129, Elizalde_2018, Mokeeva_2013, Nabulsi_2010, Ghosh_2021_36}. In an altogether different context, Jafferis \textit{et al.}~\cite{Jafferis_2022} reached a notable milestone in quantum simulation, realizing a traversable-wormhole scenario on Google's Sycamore quantum processor. Follow-up studies, however, raised doubts about the result, arguing that the experimental outcomes did not sit well with expected gravitational behavior \cite{Kobrin_2023}.

Although classical matter models forbid NEC violation, such violations arise quite naturally within many dark energy scenarios, including those built on negative-pressure fluids and phantom-like behaviour \cite{Lobo_2005, Visser_1995}. This makes GGDE, with its exotic character, a plausible candidate for sourcing wormhole geometries. Earlier studies have examined the links between dark energy and wormhole spacetimes, proposing that the required exotic behavior could arise from quantum vacuum effects or fluctuations in the energy-momentum tensor \cite{Rahaman_2007, Rahaman_2019, RAHAMAN2006}.

The present-day cosmic acceleration has, in turn, motivated a broad family of dark energy models in which the equation-of-state (EoS) parameter, $w = p/\rho$, is allowed to depart from the cosmological constant value $w = -1$. Given that the microphysical origin of this component remains unknown, it has become standard practice to describe it using phenomenological EoS parametrizations that track its possible evolution with redshift while remaining simple enough to test against data. Foremost among these is the Chevallier--Polarski--Linder (CPL) form, whose compact two-parameter structure and good behavior across the full cosmological history have made it something of a benchmark choice \cite{Chevallier2001, Linder2003, Ghosh2026de}. A number of related parametrizations -- of Jassal--Bagla--Padmanabhan type, Barboza--Alcaniz type, and Seljak type, among others \cite{Mukherjee2026de, Paramanik2026de, Basak2026de, Basak2025de} -- broaden this picture further, allowing in particular for an EoS that crosses the phantom divide at $w=-1$. Such a phantom regime, $w<-1$, matters directly for wormhole physics: a Morris--Thorne wormhole throat can only stay open by virtue of precisely this kind of null-energy-condition breakdown \cite{metric, MorrisThorneYurtsever1988}. Extending the original traversable-wormhole construction \cite{metric, MorrisThorneYurtsever1988}, several groups have obtained wormhole geometries -- static, spherically symmetric -- in which phantom-like fluids with a barotropic equation of state supply the required stress-energy \cite{Sushkov2005, Lobo2005a}, going on to study the resulting linear stability \cite{Lobo2005b} and to identify cases admitting exact solutions \cite{Zaslavskii2005}. More recently, such redshift-dependent EoS parametrizations have been folded directly into the stress-energy content of the wormhole itself, so that the shape function and redshift function inherit the same $(w_0,w_a)$-type parameters that supernova, baryon-acoustic-oscillation, and cosmic-chronometer data are used to constrain \cite{WangMeng2016}, with the same reconstruction strategy extended into holographic and other non-standard dark-energy frameworks \cite{GarattiniChannuie2023}. The present work sits at exactly this junction, where reconstructing the cosmological EoS meets the machinery used to generate wormhole solutions.

A frequently used strategy for modifying gravity is to generalize the Einstein--Hilbert action of GR, yielding models in which spacetime curvature plays an even more explicit role. Familiar examples of this kind include $f(R)$ gravity, $f(G)$ gravity (a modified Gauss--Bonnet variant), $f(R, T)$ gravity (which couples curvature to the matter sector), and $f(R,G)$ gravity (which combines the Ricci scalar with the Gauss--Bonnet invariant), among others \cite{NOJIRI_2011, Radhakrishnan_2024, Gadbail_2024, DeFelice_2010, Fayaz_2016, Singh_2016_13, Sun_2016_25, CLIFTON_20121, Chakraborty2013_45, SARKAR_2024101439, Mondal_2024, Myrzakulov_2011, Khachatryan_2011_697, Bamba_2017_49}. A different route to modifying gravity replaces curvature with torsion as the underlying geometric quantity. The resulting framework, teleparallel gravity \cite{Aldrovandi_2012, Andrade_2002}, dispenses with both the torsion-free Levi-Civita connection and the metric tensor as foundational objects. In their place, it relies on a curvature-free Weitzenb{\"o}ck connection together with a set of vierbein (tetrad) fields. This line of reasoning gives rise to the teleparallel equivalent of general relativity (TEGR) \cite{Garecki_2010, Maluf_2013, Arcos_2004}: a formulation that reproduces the same field equations as GR while admitting a rather different mathematical and physical interpretation. In particular, TEGR's dynamics resemble the force-based formulation of classical electrodynamics, in contrast to the geodesic-motion picture of standard GR. One consequence of this is that TEGR allows for departures from the weak equivalence principle, something Einstein's original theory does not permit \cite{Aldrovandi_2004_36, Aldrovandi_2004_34}.

The central aim of this study is to determine whether static, spherically symmetric wormholes can be sustained by GGDE. We begin in \cref{sec2} by reviewing the essential features and behavior of the GGDE model. Section \ref{secmet} then lays out the methodology and datasets used in the MCMC analysis. The resulting observational findings are summarized in Section \ref{sec4}. \cref{sec5} adopts a general wormhole metric and briefly outlines the basic tools needed to study wormhole geometry. \cref{sec6} presents a detailed analysis of the geometry that arises when GGDE acts as the source, deriving a set of wormhole solutions from Einstein's field equations. \cref{sec7} examines the linearized stability of the resulting thin--shell wormhole structures under radial perturbations. Our findings are then brought together and summarized in \cref{sec8}. Throughout, we work in units where $c = G = 1$ -- with $c$ the \textit{speed of light in vacuum} and $G$ the \textit{universal gravitational constant} -- to simplify the calculations; consequently, every physical quantity in this study carries the dimensions of $length$ (L) alone. The kiloparsec (KPC) is adopted throughout as the unit of length.


\section{A Review on GGDE}\label{sec2}

A new class of dark energy models, known as \textit{Ghost Dark Energy} (GDE), has recently attracted attention by introducing an additional degree of freedom, inspired by the Veneziano ghost fields emerging from quantum chromodynamics (QCD). In this framework, the vacuum energy density contributed by these ghost fields is directly proportional to $\Lambda_{\text{QCD}}^3 H$, where $\Lambda_{\text{QCD}}$ denotes the characteristic mass scale of QCD, and $H$ represents the Hubble parameter \cite{Ohta_2012}.

In flat Minkowski spacetime, these ghost fields make no contribution to vacuum energy. However, in curved spacetime backgrounds \cite{URBAN_2010135,Urban_2009_80,URBAN_20109}, the resulting energy density scales as $\sim (3 \times 10^{-3}~\text{eV})^4$, a value that aligns well with observed cosmic acceleration, thus naturally avoiding the fine-tuning problem commonly associated with the cosmological constant \cite{EBRAHIMI_2011_20,Cai_2011_84}. In its simplest form, the energy density of ghost dark energy is expressed as $\rho_D = \alpha H$, where $\alpha$ is a constant and $H = \dot{a}/a$, with $a$ being the scale factor.

Nonetheless, theoretical analyses suggest that the vacuum expectation value of the energy-momentum tensor, when conserved in isolation, implies the energy contribution from vacuum fields cannot strictly scale linearly with $H$. This leads to the emergence of an additional $H^2$ term. Consequently, the total energy density can be expanded as $H + \mathcal{O}(H^2)$ \cite{Maggiore_2011,Zhitnitsky_2012_86}. The inclusion of this subdominant quadratic term has significant implications for modelling the early dynamics of the universe \cite{Sheykhi_2012_339,Sheykhi_2012_44,Sheykhi_2011_95}. Studies have demonstrated that this extension improves consistency with observational data when compared to the original GDE model. As a result, this modified version, incorporating both linear and quadratic terms in $H$, is referred to as the \textit{Generalized Ghost Dark Energy} (GGDE) model. In this formulation, the energy density of the generalised ghost dark energy (GGDE) takes the following form \cite{Biswas_2019_79, Cai_2011_84, EBRAHIMI_2011_20, Biswas_2022_37}:

\begin{equation}\label{eq1}
    \rho_{D} ~=~ \frac{3}{8\pi} (\alpha H + \beta H^2) ~.
\end{equation}

Here, $\alpha$ and $\beta$ are constants and $H \left( =\frac{\dot{a}}{a} \right)$ is the Hubble parameter. It should be noted that here the term $\frac{3}{8\pi}$ is multiplied to make calculations easier.

The Friedmann equation for a flat universe is given by \cite{Wang_2016_76}:

\begin{equation}\label{eq2}
    H^2 ~=~ \frac{8 \pi}{3} \sum_i \rho_i ~.
\end{equation}

The summation runs over the energy densities for non-relativistic matter and dark energy components.

By setting $x ~=~ \log a$, we can rewrite \cref{eq2} as follows:

\begin{equation}\label{eq3}
    H^2 ~=~ \frac{8 \pi}{3} \left\{\frac{3}{8 \pi}(\alpha H + \beta H^2) + \rho_m e^{-3 x} \right\} ~, 
\end{equation}
where $\rho_m$ denotes the energy density due to non-relativistic matter. We now introduce $h$ as the scaled Hubble constant by taking $h = \frac{H}{H_0}$ and thus, \cref{eq3} becomes:

\begin{equation}\label{eq4}
    h^2 ~=~ \left( \frac{\alpha h}{H_0} + \beta h^2 \right) + \Omega_{m0} e^{-3 x} ~, 
\end{equation}
where $\Omega_{m0} = \frac{8 \pi \rho_m}{3 H_0^2}$ is the relative density of non-relativistic matter 

Solving \cref{eq4} for $h$ we get:

\begin{equation}\label{eq5}
    h_{\pm} ~=~ \frac{- \alpha \pm \sqrt{\alpha ^2-4 H_0^2 (\beta -1) e^{-3 x} \Omega _{\text{m0}}}}{2 H_0 (\beta -1)} ~.
\end{equation}

From \cref{eq5}, we get two values of $h$, namely $h_+$ and $h_-$, but we only take $h_+$ as the value of $h$ because it results in the universe's expansion, whereas $h_-$ leads to a shrinking universe. Also, from this point onward, we shall write $h$ instead of $h_+$ for simplicity. Hence, \cref{eq5} gives:

\begin{equation}\label{eq6}
    h ~=~ \frac{- \alpha + \sqrt{\alpha ^2-4 H_0^2 (\beta -1) e^{-3 x} \Omega _{\text{m0}}}}{2 H_0 (\beta -1)} ~.
\end{equation}

To relate our results with various observational data, we now set $z ~=~ \frac{1}{a} - 1$, where $z$ denotes the redshift. Thus from \cref{eq6} we get:

\begin{equation}\label{eq7}
    h^2(z) ~=~ \frac{\left\{\sqrt{\alpha ^2-4 H_0^2 (\beta -1) (z+1)^3 \Omega _{\text{m0}}} - \alpha\right\}^2}{4 H_0^2 (\beta -1)^2} ~.
\end{equation}

Thus, the energy density of GGDE from \cref{eq1} turns out to be:

\begin{equation}\label{eq8}
\begin{aligned}
    \rho_D ~=~ \frac{3 \left\{\alpha ^2-\alpha \sqrt{\alpha ^2 - 4 H_0^2 (\beta -1) e^{-3 x} \Omega _{\text{m0}}}-2 (\beta -1) \beta  H_0^2 e^{-3 x} \Omega _{\text{m0}}\right\}}{16 \pi  (\beta -1)^2} ~.
\end{aligned}
\end{equation}

If we assume that the energy in this system is conserved, we get:

\begin{equation}\label{eq9}
    \frac{d\rho}{dx} + 3 (\rho +p) ~=~ 0 ~,
\end{equation}
where $p$ denotes the pressure of the system. Now using \cref{eq8} we get the pressure for GGDE as:

\begin{equation}\label{eq10}
\begin{aligned}
    p_D ~=~ \frac{3 \alpha  e^{-3 x} \left\{-2 (\beta -1) H_0^2 \Omega _{\text{m0}} - \alpha e^{3 x} \sqrt{\alpha ^2 - 4  H_0^2 (\beta -1) e^{-3 x} \Omega _{\text{m0}}}+\alpha ^2 e^{3 x}\right\}}{16 \pi  (\beta -1)^2 \sqrt{\alpha ^2 - 4 H_0^2 (\beta -1) e^{-3 x} \Omega _{\text{m0}}}} ~.
\end{aligned}
\end{equation}

If we substitute the relation $z ~=~ \frac{1}{a} - 1$ in both \cref{eq8} and \cref{eq10}, we get the following two modified equations:

\begin{equation}\label{eq11}
\begin{aligned}
    \rho_D ~=~ \frac{3 \left\{\alpha ^2-\alpha \sqrt{\alpha ^2 - 4 H_0^2 (\beta -1) (z+1)^3 \Omega _{\text{m0}}}-2 (\beta -1) \beta  H_0^2 (z+1)^3 \Omega _{\text{m0}}\right\}}{16 \pi  (\beta -1)^2} ~,
\end{aligned}
\end{equation}

\begin{equation}\label{eq12}
\begin{aligned}
    p_D ~=~ \frac{3 \alpha \left\{\alpha^2-2 (z+1)^3 (\beta -1) H_0^2 \Omega _{\text{m0}} - \alpha \sqrt{\alpha ^2 - 4  H_0^2 (\beta -1) (z+1)^3 \Omega _{\text{m0}}}\right\}}{16 \pi  (\beta -1)^2 \sqrt{\alpha ^2-4 H_0^2 (\beta -1) (z+1)^3 \Omega _{\text{m0}}}} ~.
\end{aligned}
\end{equation}


\section{Methodology and Datasets}\label{secmet}
To constrain the free parameters of the proposed cosmological model and examine its observational viability, we compare its theoretical predictions with three independent probes of the cosmic expansion history spanning a wide range of redshifts. These include the Cosmic Chronometer (CC) measurements, the Baryon Acoustic Oscillation (BAO) observations, and the $Pantheon^+$ Type Ia Supernova (SNe Ia) compilation. Each dataset probes the expansion history through a different physical mechanism: CC data provide direct and nearly model-independent measurements of the Hubble parameter, BAO observations serve as a standard ruler for constraining cosmic distances, while $Pantheon^+$ SNe Ia act as standard candles that accurately trace the late-time expansion of the Universe. Their combination significantly reduces parameter degeneracies and systematic uncertainties, leading to more robust constraints on the model parameters. The statistical framework adopted for each dataset is described in the following subsections.

\subsection{MCMC-Based Bayesian Parameter Estimation}

The parameter vector
\[
\Theta=(H_0,\Omega_{m0},\alpha,\beta)
\]
is constrained within the Bayesian inference framework \cite{Trotta:2017wnx,akeret2013cosmohammer}. The posterior probability distributions are explored using the Markov Chain Monte Carlo (MCMC) technique \cite{Foreman-Mackey:2012any}, which provides an efficient sampling of the multidimensional parameter space while simultaneously estimating parameter uncertainties and their mutual correlations.

According to Bayes' theorem, the posterior probability distribution is expressed as
\begin{equation}
P(\Theta|D,I)=
\frac{P(\Theta|I)\,P(D|\Theta,I)}
{P(D|I)},
\end{equation}
where $P(\Theta|I)$ denotes the prior distribution of the model parameters, $P(D|\Theta,I)$ represents the likelihood function for the observational data $D$, and $P(D|I)$ is the Bayesian evidence. Since the evidence is independent of the model parameters, it acts only as a normalization constant and is not required during parameter estimation.

Assuming Gaussian-distributed observational uncertainties, the likelihood function is written as
\begin{equation}
\mathcal{L}(\Theta)=
\exp\left[-\frac{\chi^2(\Theta)}{2}\right].
\label{eq:likelihood}
\end{equation}

Because the CC, BAO, and $Pantheon^+$ datasets are treated as statistically independent, the total chi-squared function is given by
\begin{equation}
\chi^2_{\rm total}(\Theta)
=
\chi^2_{\rm CC}(\Theta)
+
\chi^2_{\rm BAO}(\Theta)
+
\chi^2_{\rm Pantheon+}(\Theta).
\label{eq:chi_total}
\end{equation}

The individual likelihood functions corresponding to each observational probe are presented below.

\subsection{Cosmic Chronometer Data}

The Cosmic Chronometer (CC) method offers one of the few direct and nearly model-independent measurements of the Hubble expansion rate. Rather than relying on an assumed cosmological model, this technique estimates $H(z)$ by measuring the differential age evolution of massive, passively evolving galaxies. The Hubble parameter is obtained through
\begin{equation}
H(z)=
-\frac{1}{1+z}\frac{dz}{dt},
\end{equation}
where the quantity $dz/dt$ is determined from the age difference between galaxies separated by a small redshift interval. Since these galaxies experience minimal subsequent star formation after their initial formation epoch, they function as reliable cosmic clocks, making the CC method an important and complementary probe of the expansion history.

In the present analysis, we employ a compilation of 32 CC measurements \cite{Moresco2012, Moresco2015, Moresco2016, Zhang2014, Ratsimbazafy2017} covering the redshift interval
\[
0.07 \leq z \leq 1.965.
\]
These measurements are obtained from several independent spectroscopic galaxy surveys and provide direct constraints on the expansion rate over intermediate redshifts. The corresponding chi-squared statistic is defined as
\begin{equation}
\chi^2_{\rm CC}(\Theta)
=
\sum_{i=1}^{32}
\frac{\left[H_{\rm th}(z_i;\Theta)-H_{\rm obs}(z_i)\right]^2}
{\sigma_H^2(z_i)},
\label{eq:chi_cc}
\end{equation}
where $H_{\rm th}(z_i;\Theta)$ and $H_{\rm obs}(z_i)$ denote the theoretical and observed Hubble parameters at redshift $z_i$, respectively, while $\sigma_H(z_i)$ represents the corresponding observational uncertainty.

\subsection{$Pantheon^+$ Type Ia Supernovae}

Type Ia supernovae (SNe Ia) are well-established standardizable candles that provide precise measurements of the luminosity distance over a broad range of redshifts, making them one of the most powerful observational probes of the late-time expansion history of the Universe. In this work, we employ the $Pantheon^+$ compilation \cite{Scolnic2022, Brout2022, Riess2022}, which is currently the largest and most comprehensive SNe Ia dataset available. It consists of 1701 light curves corresponding to 1550 spectroscopically confirmed SNe Ia distributed over the redshift interval
\[
0.00122 \leq z \leq 2.2613.
\]
The compilation combines observations from several ground-based surveys together with measurements from the \textit{Hubble Space Telescope} (HST), thereby extending both the redshift coverage and statistical power beyond the original Pantheon sample. In addition, $Pantheon^+$ incorporates significant improvements in photometric calibration, zero-point determination, host-galaxy mass corrections, peculiar-velocity modeling, and the treatment of redshift-dependent systematic uncertainties, resulting in more accurate and robust luminosity distance measurements.

For a given cosmological model, the theoretical distance modulus is expressed as
\begin{equation}
\mu_{\rm th}(z;\Theta)
=
5\log_{10}\!\left[d_L(z;\Theta)\right]
+25,
\label{eq:mu_theory}
\end{equation}
where the luminosity distance is given by
\begin{equation}
d_L(z)
=
(1+z)c
\int_{0}^{z}
\frac{dz'}{H(z';\Theta)},
\label{eq:luminosity_distance}
\end{equation}
with $c$ denoting the speed of light and $z'$ representing the integration variable.

Unlike datasets with independent measurements, the $Pantheon^+$ sample exhibits non-negligible correlations among different supernova observations owing to common sources of systematic uncertainty, including photometric calibration, zero-point offsets, survey selection effects, and other instrumental and astrophysical systematics. Consequently, the full covariance matrix is employed in the likelihood analysis to consistently account for both statistical and systematic uncertainties.

The corresponding chi-squared statistic is therefore written as
\begin{equation}
\chi^2_{\rm Pantheon+}(\Theta)
=
\Delta\boldsymbol{\mu}^{\,T}
\mathbf{C}^{-1}
\Delta\boldsymbol{\mu},
\label{eq:chi_pantheon}
\end{equation}
where $\mathbf{C}$ denotes the full covariance matrix and
\begin{equation}
\Delta\mu_i
=
\mu_{\rm obs}(z_i)
-
\mu_{\rm th}(z_i;\Theta)
\label{eq:delta_mu}
\end{equation}
is the residual between the observed and theoretical distance moduli for the $i$-th supernova. The analysis includes all 1701 measurements contained in the $Pantheon^+$ compilation. Incorporating the complete covariance matrix ensures that correlations between supernova observations are properly propagated into the parameter estimation, thereby yielding statistically reliable confidence intervals and preventing artificially stringent constraints that could arise from assuming uncorrelated measurement errors.
\subsection{Baryon Acoustic Oscillations (BAO)}

Baryon Acoustic Oscillations (BAO) arise from pressure-driven sound waves that propagated through the tightly coupled baryon-photon plasma during the early Universe \cite{Ross2015, Alam2021, Alam2017,DESI2024, Beutler2011}. Prior to the epoch of recombination, the competition between gravitational attraction and radiation pressure generated acoustic oscillations in the primordial plasma. Following the drag epoch ($z_d$), when baryons decoupled from photons, these oscillations became imprinted on the large-scale matter distribution, leaving a characteristic comoving scale known as the sound horizon. This scale is given by
\begin{equation}
r_d=
\int_{z_d}^{\infty}
\frac{c_s(z)}{H(z)}\,dz,
\label{eq:rd}
\end{equation}
where the sound speed of the baryon--photon fluid is
\begin{equation}
c_s(z)=
\frac{c}
{\sqrt{3\left[1+\dfrac{3\Omega_b}
{4\Omega_\gamma(1+z)}\right]}},
\label{eq:cs}
\end{equation}
with $\Omega_b$ and $\Omega_\gamma$ representing the present-day baryon and photon density parameters, respectively. The sound horizon has a characteristic scale of approximately $147\,\mathrm{Mpc}$ and serves as a standard ruler for measuring the cosmic expansion history through the large-scale clustering of galaxies.

BAO observations constrain different cosmological distance measures depending on the orientation of the acoustic feature relative to the observer. The transverse BAO signal is characterized by the comoving angular diameter distance,
\begin{equation}
D_M(z)=
\frac{c}{H_0}
\int_0^z
\frac{dz'}{E(z')},
\label{eq:DM}
\end{equation}
where $E(z)=H(z)/H_0$ is the normalized Hubble parameter. Along the line of sight, the relevant quantity is the Hubble distance,
\begin{equation}
D_H(z)=
\frac{c}{H(z)},
\label{eq:DH}
\end{equation}
while isotropic BAO analyses commonly employ the volume-averaged distance,
\begin{equation}
D_V(z)=
\left[
z\,D_M^2(z)\,D_H(z)
\right]^{1/3}.
\label{eq:DV}
\end{equation}

In observational analyses, these distance measures are generally reported in units of the sound horizon through the dimensionless ratios $D_M/r_d$, $D_H/r_d$, and $D_V/r_d$. Since BAO measurements rely on a geometric standard ruler and are affected by different systematic uncertainties than Cosmic Chronometer and Type Ia Supernova observations, they provide highly complementary information. Consequently, their combination with CC and $Pantheon^+$ datasets substantially improves the precision of cosmological parameter estimation while helping to reduce parameter degeneracies.

\section{Findings}\label{sec4}
\begin{figure}[h]
    \centering
    \begin{subfigure}[b]{0.45\textwidth}
        \includegraphics[width=\textwidth]{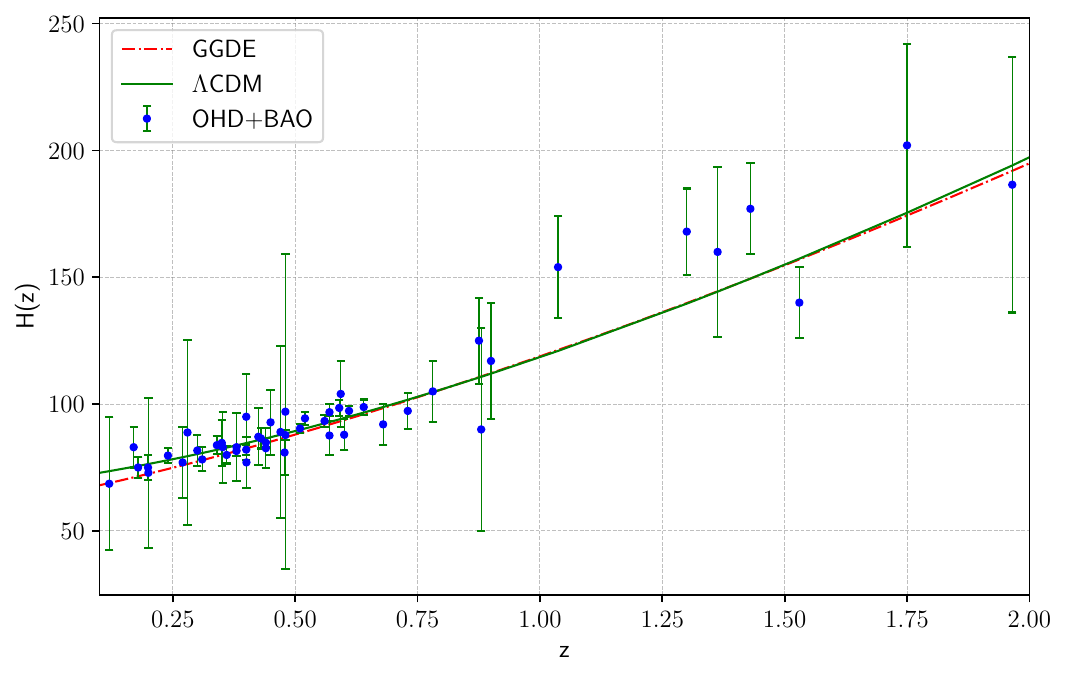}
        \caption{Comparison of Hubble Parameter with $\Lambda$CDM using 57 CC + BAO data points.}
        \label{fig:Hubble}
    \end{subfigure}
    \begin{subfigure}[b]{0.45\textwidth}
        \includegraphics[width=\textwidth]{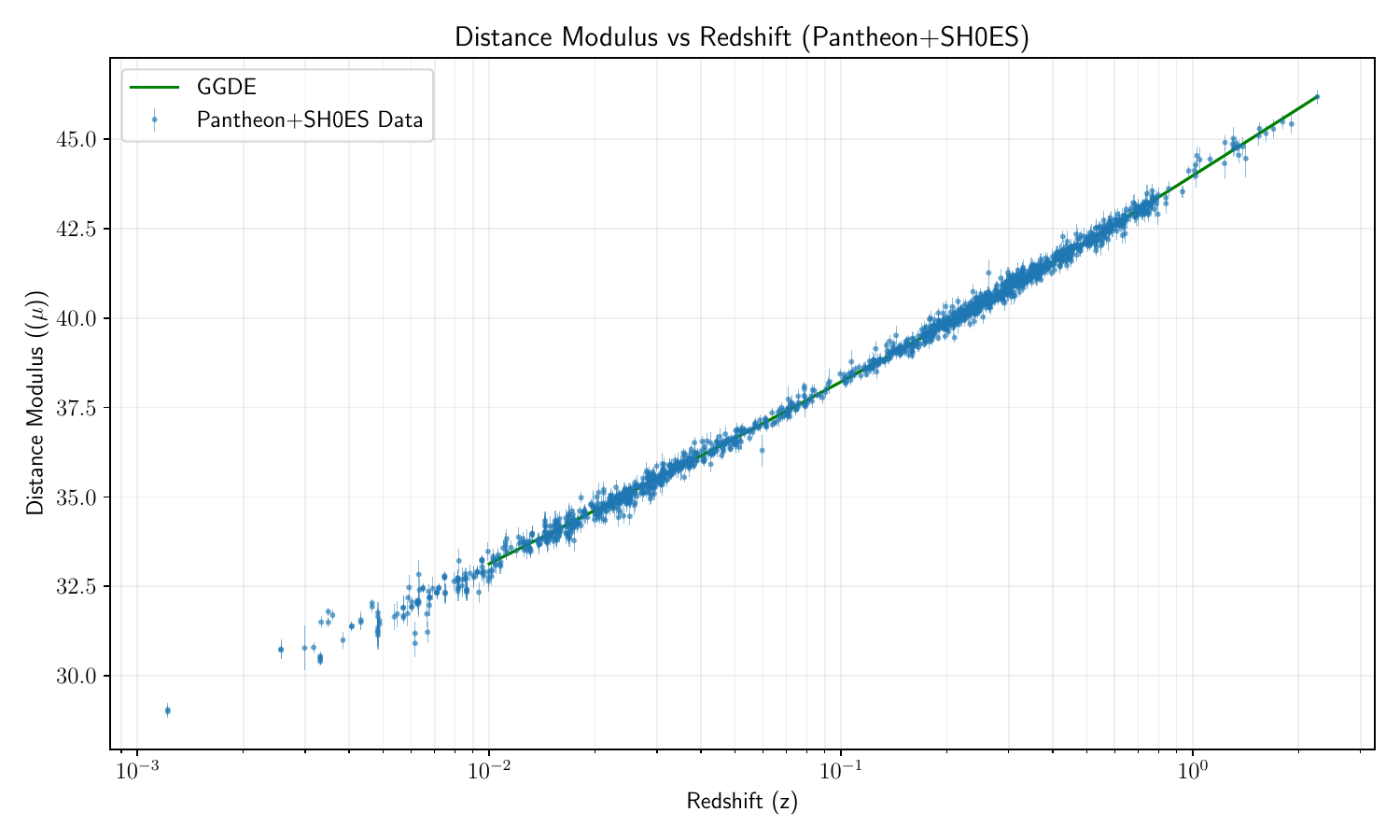}
        \caption{Plot of Distance Modulus using 1701 $Pantheon^{+}$ data points.}
        \label{fig:distance modulus}
    \end{subfigure}
    \caption{}
\end{figure}

As illustrated in Fig.\ref{fig:Hubble}, the reconstructed Hubble parameter predicted by the proposed model is in excellent agreement with the observational data throughout the considered redshift range. The theoretical evolution closely follows the corresponding prediction of the standard $\Lambda$CDM model, indicating that the present model successfully reproduces the observed expansion history of the Universe. Fig.~\ref{fig:distance modulus} presents the reconstructed distance modulus obtained using the best-fit parameters of the proposed model together with the $Pantheon^+$ Type Ia supernova dataset. It is evident that the theoretical distance modulus provides an excellent fit to the observed supernova measurements over the entire redshift range, confirming that the model accurately reproduces the luminosity distance-redshift relation. The overall agreement between the theoretical predictions and the observational datasets demonstrates that the proposed model offers a viable description of the late-time expansion of the Universe while remaining fully consistent with current cosmological observations.
\begin{table*}[t]
\centering
\renewcommand{\arraystretch}{4} 
\setlength{\tabcolsep}{10pt} 
\caption{Best-fit parameters obtained from Pantheon+SH0ES, OHD+BAO, and Joint datasets.}
\begin{tabular}{|c|c|c|c|c|}
\hline
\textbf{Dataset} & \boldmath{$H_{0}$} & \boldmath{$\Omega_{m0}$} & \boldmath{$\alpha$} & \boldmath{$\beta$} \\ 
\hline
Pantheon+SH0ES & $72.483^{+2.213}_{-2.489}$ & $0.278^{+0.016}_{-0.017}$ & $0.183^{+0.022}_{-0.017}$ & $0.786^{+0.019}_{-0.020}$ \\ 
\hline
OHD+BAO & $72.236^{+2.686}_{-2.510}$ & $0.276^{+0.014}_{-0.015}$ & $0.174^{+0.016}_{-0.015}$ & $0.765^{+0.016}_{-0.015}$ \\ 
\hline
Joint & $72.446^{+2.430}_{-2.336}$ & $0.278^{+0.015}_{-0.020}$ & $0.265^{+0.020}_{-0.016}$ & $0.715^{+0.018}_{-0.020}$ \\ 
\hline
\end{tabular}
\label{tab:bestfit_parameters}
\end{table*}

\begin{figure}[htbp]
    \centering
    \begin{subfigure}[b]{0.7\textwidth}
        \includegraphics[width=\textwidth]{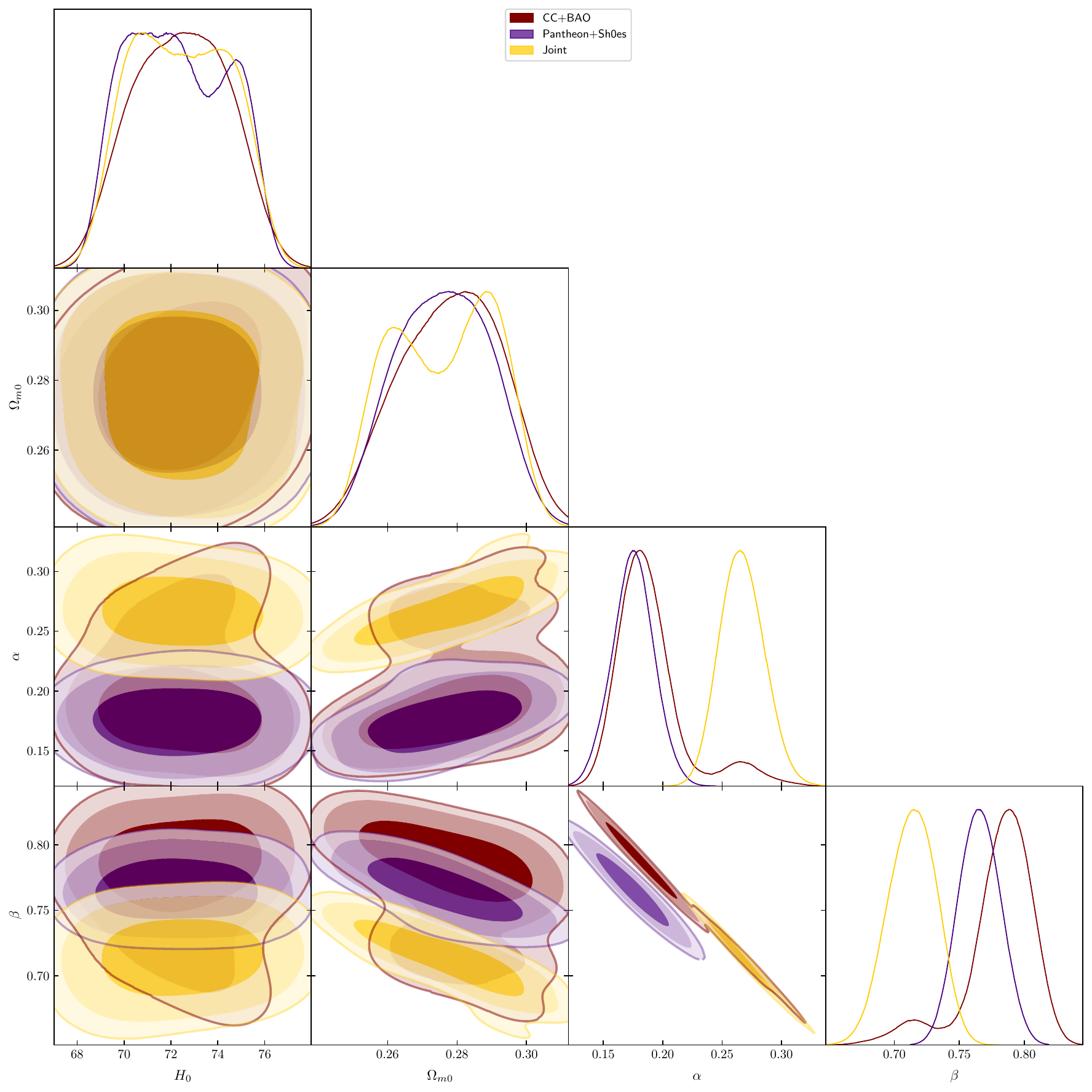}
    \end{subfigure}
    \caption{Posterior distribution of model parameters at $1\sigma$ and $2\sigma$ confidence levels.}
    \label{fig:getdist}
\end{figure}
Figure~\ref{fig:getdist} shows the marginalized one-dimensional posterior distributions and the corresponding two-dimensional confidence contours for the model parameters obtained from the CC+BAO, $Pantheon^+$, and the combined CC+BAO+$Pantheon^+$ datasets. The contours represent the $1\sigma$ (68\%) and $2\sigma$ (95\%) confidence intervals. It is evident that the combination of these complementary observations effectively reduces parameter degeneracies and provides more precise and robust constraints on the model parameters. The corresponding best-fit values and their $1\sigma$ uncertainties are summarized in Table~\ref{tab:bestfit_parameters}. The Hubble constant, $H_0$, is consistently constrained across all three datasets, with best-fit values lying in the narrow interval $72.24$-$72.48~\mathrm{km\,s^{-1}\, Mpc^{-1}}$, indicating that the inferred value of $H_0$ is largely insensitive to the choice of observational probe. Similarly, the present matter density parameter, $\Omega_{m0}$, exhibits remarkable stability among the three dataset combinations, with best-fit values clustered around $0.28$. The close agreement of these estimates demonstrates the robustness of the model in reproducing the observed matter content of the Universe and is consistent with the expectations of the standard cosmological scenario.

In comparison, the model-specific parameters $\alpha$ and $\beta$ display more noticeable variations between the individual and joint analyses. The combined dataset yields tighter constraints on both parameters, with the posterior distributions becoming significantly narrower than those obtained from the individual datasets. This improvement indicates that the complementary nature of the CC+BAO and $Pantheon^+$ observations effectively breaks the parameter degeneracies present in the separate analyses, leading to more precise estimates of the model parameters.

Overall, the high degree of consistency in the estimates of $H_0$ and $\Omega_{m0}$, together with the improved constraints on $\alpha$ and $\beta$ obtained from the joint analysis, demonstrates the statistical robustness of the proposed model.\\

To further assess the performance of the proposed cosmological model, we employ the Akaike Information Criterion (AIC) and the Bayesian Information Criterion (BIC), both of which provide quantitative measures for model comparison by balancing the goodness of fit against the complexity of the model.

The AIC and BIC are defined as
\begin{equation}
\mathrm{AIC}=\chi_{\min}^2+2k,
\end{equation}
and
\begin{equation}
\mathrm{BIC}=\chi_{\min}^2+k\ln N,
\end{equation}
where $k$ denotes the number of free parameters and $N$ is the total number of observational data points. 

The values of AIC and BIC obtained for the CC+BAO, $Pantheon^+$, and the combined $CC+BAO+Pantheon^+$ analyses are summarized in Table~\ref{tab:model_metrics_comparison}. For the CC+BAO dataset, $\Lambda$CDM is favored, having lower AIC and BIC values. While the small $\Delta$AIC ($=1.13$) suggests that both models provide comparable fits. In contrast, the proposed model is preferred for both the Pantheon$^{+}$ and joint datasets, yielding lower AIC and BIC values than $\Lambda$CDM. For the Pantheon$^{+}$ dataset, the differences ($\Delta$AIC $=4.95$ and $\Delta$BIC $=3.87$) indicate substantial evidence supporting the proposed model. For the joint dataset, the preference remains in favor of the proposed model, although the smaller differences ($\Delta$AIC $=1.62$ and $\Delta$BIC $=0.68$) suggest both models provide similarly good fits.\\
\begin{table}[ht]
\centering
\caption{AIC, BIC, and their corresponding differences ($\Delta$AIC and $\Delta$BIC) for the proposed model and the $\Lambda$CDM model using different observational datasets.}
\label{tab:model_metrics_comparison}
\begin{tabular}{|l|c|c|c|c|}
\hline
\textbf{Model} & \textbf{AIC} & \textbf{BIC} & $\mathbf{\Delta}$\textbf{AIC} & $\mathbf{\Delta}$\textbf{BIC} \\
\hline

\multicolumn{5}{|c|}{\textbf{CC+BAO Dataset}} \\
\hline
Model & 37.37 & 45.54 & 1.13 & 5.22 \\
$\Lambda$CDM & 36.24 & 40.32 & 0.00 & 0.00 \\
\hline
\multicolumn{5}{|c|}{\textbf{Pantheon$^{+}$ Dataset}} \\
\hline
Model & 1702.08 & 1709.84 & 0.00 & 0.00 \\
$\Lambda$CDM & 1707.03 & 1713.71 & 4.95 & 3.87 \\
\hline
\multicolumn{5}{|c|}{\textbf{Joint}} \\
\hline
Model & 1824.48 & 1836.37 & 0.00 & 0.00 \\
$\Lambda$CDM & 1826.10 & 1837.05 & 1.62 & 0.68 \\
\hline
\end{tabular}
\end{table}
\begin{table}[t]
\centering
\renewcommand{\arraystretch}{1.3} 
\setlength{\tabcolsep}{10pt} 
\caption{Gelman-Rubin convergence diagnostic (\(\hat{R}\)) values for different model parameters using Pantheon+SH0ES (SN), OHD+BAO, and Joint datasets.}
\begin{tabular}{|c|c|c|c|}
\hline
\textbf{Parameter} & \textbf{SN} & \textbf{OHD+BAO} & \textbf{Joint} \\ 
\hline
\boldmath{$H_{0}$} & 1.030 & 1.005 & 0.999 \\ 
\hline
\boldmath{$\Omega_{m0}$} & 1.055 & 0.992 & 0.995 \\ 
\hline
\boldmath{$\alpha$} & 2.031 & 0.989 & 1.011 \\ 
\hline
\boldmath{$\beta$} & 1.886 & 0.984 & 1.007 \\ 
\hline
\end{tabular}
\label{tab:rhat_values}
\end{table}
The Gelman-Rubin convergence diagnostic ($\hat{R}$) for each free parameter obtained from the CC+BAO, $Pantheon^+$, and the combined CC+BAO+$Pantheon^+$ analyses is presented in Table~\ref{tab:rhat_values}. For both the CC+BAO and the joint datasets, the $\hat{R}$ values lie very close to unity, ranging from 0.984 to 1.011, thereby satisfying the commonly adopted convergence criterion ($\hat{R}<1.1$). This indicates that the independent MCMC chains have converged successfully and provide reliable sampling of the posterior distributions. In contrast, the $Pantheon^+$ analysis exhibits comparatively weaker convergence. Although the parameters $H_0$ ($\hat{R}=1.030$) and $\Omega_{m0}$ ($\hat{R}=1.055$) satisfy the convergence criterion, the parameters $\alpha$ and $\beta$ show relatively large values of $\hat{R}=2.031$ and $\hat{R}=1.886$, respectively. These elevated values suggest incomplete chain mixing and indicate that the $Pantheon^+$ dataset alone provides comparatively weaker constraints on these model parameters, resulting in broader posterior distributions and residual parameter degeneracies.

The inclusion of the CC+BAO observations in the joint analysis substantially improves the convergence behavior by reducing these degeneracies. Consequently, all model parameters exhibit well-converged $\hat{R}$ values within the interval $0.999$-$1.011$, demonstrating the statistical robustness of the combined analysis and confirming that the joint dataset provides the most reliable parameter constraints obtained in the present work.


\section{Basics of Wormhole}\label{sec5}

\subsection{Wormhole geometry}

The following line element represents the spherically symmetric, static wormhole geometry \cite{metric}:

\begin{equation}\label{eq4.1}
    ds^{2} ~=~ - e^{2\Phi(r)} dt^2 + \frac{dr^2}{1 - \frac{b(r)}{r}} + r^2 (d\theta^2 + \sin^2 \theta d\phi^2) ~,
\end{equation}
where the functions $\Phi(r)$ and $b(r)$ are known as \textit{redshift function} and \textit{shape function}, respectively. All these functions depend upon $r$, the radial coordinate, which runs over the values $[r_0,\infty)$, $r_0$ being known as the radius of the wormhole throat, and throughout this article, we have considered $r_0=1$. Also, we shall be considering a cut-off radius, denoted by `$a$', of the stress-energy tensor $T_{\mu\nu}$ wherever necessary. In \cref{eq4.1}, $\theta$ and $\Phi$ denote the angular coordinates and they satisfy $0 \leq \theta \leq \pi$ and $0 \leq \Phi \leq 2\pi$ respectively.\par

If the line element \cref{eq4.1} satisfies the following traversability conditions, the underlying wormhole becomes traversable \cite{PARAMANIK_2025}:

\begin{enumerate}

    \item \textbf{Throat Condition:} $b(r_{0}) = r_{0} ~\text{and}~ 1 - \frac{b(r)}{r} > 0 ~\forall~ r \in \left( r_0, \infty \right)$.
    
    \item \textbf{Flaring-out condition:} $b'(r_{0}) ~<~ 1$ at $r ~=~ r_{0}$ and $b(r) - r b'(r) ~>~ 0 ~\forall~ r ~\in~ \left( r_0, \infty \right)$.
    
    \item \textbf{Asymptotic flatness condition:} $\lim_{r\rightarrow \infty} \frac{b(r)}{r} ~=~ 0$. This condition tells us that the traversable wormhole spacetime eventually becomes flat.
    
    \item \textbf{No horizon condition:} The redshift function $\Phi(r)$ must be bounded and continuous all over the range considered. 
    This condition ensures that there is no kind of horizon in the traversable wormhole spacetime.
    
\end{enumerate}

We now implement the Einstein field equations $G_{\mu\nu} ~=~ T_{\mu\nu}$ to obtain the following equations relating $b(r)$ with $\Phi(r)$ as:

\begin{align}
    b' ~=~& r^2 \rho ~, \label{eq4.2}\\
    \Phi' ~=~& \frac{b + r^3 p}{2 r^2 \left(1 - \frac{b}{r}\right)} ~, \label{eq4.3}\\
    p ~=~& \left(1 - \frac{b}{r}\right) \left\{\Phi'' + (\Phi')^2 + \frac{\Phi'}{r} - \frac{rb' - b}{2 r^2 (r-b)} - \frac{rb' - b}{2 r (r-b)} \Phi' \right\} ~.\label{eq4.4}
\end{align}

In these three equations, $\rho$ denotes the energy density, whereas $p$ denotes the pressure of the cosmic fluid. As is evident, both $\rho$ and $p$ are functions of the radial coordinate $r$ only.




Now, considering $\rho_D = \rho$ and $p_D = p$, we get the equation of state of the GGDE model by comparing \cref{eq8} and \cref{eq10} as follows:

\begin{equation}\label{eq4.6}
\begin{aligned}
    p ~=~ \frac{3 \alpha  H_0^2 e^{-3 x} \Omega _{\text{m0}}}{8 \pi  \sqrt{{\alpha ^2}-4 (\beta -1) H_0^2 e^{-3 x} \Omega _{\text{m0}}}} + \frac{\alpha \rho}{\sqrt{{\alpha ^2}-4 (\beta -1) H_0^2 e^{-3 x} \Omega _{\text{m0}}}} ~.
\end{aligned}
\end{equation}

We further introduce $\gamma ~=~ (\beta -1) H_0^2 e^{-3 x} \Omega _{\text{m0}}$ to simplify \cref{eq4.6} and thus, we get:

\begin{equation}\label{eq4.7}
    p ~=~ \frac{3 \alpha  \gamma }{8 \pi  (\beta -1) \sqrt{\alpha ^2-4 \gamma }}+\frac{\alpha \rho}{\sqrt{\alpha ^2-4 \gamma }} ~.
\end{equation}

Eliminating $\rho$ and $p$ between \crefrange{eq4.2}{eq4.3} and \cref{eq4.7} we arrive at the following equation:

\begin{equation}\label{eq4.8}
    \Phi' ~=~ \frac{r^3 \left(\frac{3 \alpha  \gamma }{8 \pi (\beta - 1)  \sqrt{\alpha ^2-4 \gamma }} + \frac{\alpha  b'}{r^2 \sqrt{\alpha ^2-4 \gamma }}\right)+b}{2 r^2 \left(1-\frac{b}{r}\right)}
\end{equation}

Any type of solution of the line element \cref{eq4.1} satisfying \cref{eq4.8} can be termed as a \textit{GGDE wormhole} solution. Also, if the solution satisfies the traversability conditions described earlier in this section, then that solution can be named as \textit{GGDE traversable wormhole} solution.

\subsection{Thin--shell Formalism}

A physically viable traversable wormhole configuration is typically required to satisfy asymptotic flatness, which is ensured if the metric functions obey the limiting behavior
\begin{equation*}
\frac{b(r)}{r} \to 0, \quad \Phi(r) \to 0 \quad \text{as} \quad r \to \infty.
\end{equation*}
In general, obtaining exact global solutions that fulfill these conditions remains a nontrivial task. A widely used alternative is to construct a composite spacetime by matching an interior wormhole geometry to an exterior vacuum solution across a timelike hypersurface. This procedure allows one to retain the desired local properties of the wormhole while ensuring appropriate asymptotic behavior. This method is widely known as the cut-and-paste technique \cite{cutpaste1, cutpaste2}.

The matching of two distinct manifolds is governed by the Darmois--Israel junction conditions \cite{Israel1966, Darmois1927, Lanczos_1924379, Sen_1924378}. If the induced metric and extrinsic curvature are continuous across the junction, the hypersurface represents a boundary surface without any localized matter content. However, if there exists a discontinuity in the extrinsic curvature, a non-vanishing surface stress-energy tensor arises, and the junction is interpreted as a thin shell \cite{Poisson2004, Visser_1995, Lobo_200571, Lobo_2005712}. In this framework, all exotic matter required to support the wormhole can be confined to the junction surface, thereby simplifying the overall physical description.

For the exterior region, we consider the Reissner--Nordstr{\"o}m spacetime, given by:

\begin{equation}\label{eq5.1}
ds^2 = -\left(1 - \frac{2M}{r} + \frac{Q^2}{r^2}\right) dt^2 
+ \left(1 - \frac{2M}{r} + \frac{Q^2}{r^2}\right)^{-1} dr^2 
+ r^2 \left(d\theta^2 + \sin^2\theta\, d\phi^2\right),
\end{equation}
where $M$ and $Q$ denote the mass and electric charge, respectively. The horizon structure of this spacetime depends on the relative magnitude of these parameters. For $|Q| < M$, two distinct horizons are present at:

\begin{equation}\label{eq5.2}
r_{\pm} = M \pm \sqrt{M^2 - Q^2},
\end{equation}
while in the extremal case ($|Q| = M$), these horizons coincide. When $|Q| > M$, no horizons exist, and the solution corresponds to a naked singularity. In order to ensure traversability and avoid event horizons in the constructed geometry, the junction radius $y$ must satisfy $y > r_+$ when $|Q| \leq M$. In contrast, for $|Q| > M$, the absence of horizons allows any choice with $y > 0$, which naturally eliminates horizon-related constraints.

The surface stress-energy tensor associated with the junction hypersurface is determined using the Lanczos equations,

\begin{equation}\label{eq5.3}
S^i_j = -\frac{1}{8\pi} \left( [K^i_j] - \delta^i_j [K] \right),
\end{equation}
where $[K^i_j]$ denotes the discontinuity in the extrinsic curvature across the shell. For a spherically symmetric configuration, the surface stress-energy tensor takes the diagonal form:

\begin{equation}\label{eq5.4}
S^i_j = \text{diag}\left(-\sigma, P, P\right),
\end{equation}
with $\sigma$ representing the surface energy density and $P$ the tangential surface pressure. If $P > 0$, then it is called as tangential surface pressure whereas if $P < 0$, then it is known as tangential surface tension.

Following the standard thin--shell formalism \cite{Poisson2004, Visser1989}, the dynamical expressions for the surface stresses of a shell located at radius $y(\tau)$ are obtained as:

\begin{equation}\label{eq5.5}
\sigma = -\frac{1}{4\pi y} \left[ 
\sqrt{1 - \frac{2M}{y} + \frac{Q^2}{y^2} + \dot{y}^2} 
- \sqrt{1 - \frac{b(y)}{y} + \dot{y}^2}
\right],
\end{equation}

\begin{align}\label{eq5.6}
P = \frac{1}{8\pi y} \left[\frac{1 - \frac{M}{y} + \dot{y}^2 + y\ddot{y}} {\sqrt{1 - \frac{2M}{y} + \frac{Q^2}{y^2} + \dot{y}^2}} - \frac{\{1 + y \Phi'(y)\}\left(1 - \frac{b(y)}{y} + \dot{y}^2\right) + y\ddot{y} - \dot{y}^2 \frac{b - y b'}{2(y - b)}}{\sqrt{1 - \frac{b(y)}{y} + \dot{y}^2}}\right],
\end{align}
where derivatives with respect to the proper time $\tau$ are denoted by overdots, and primes indicate differentiation with respect to the radial coordinate.

In the static limit of $y$ at $y_0$, characterized by $\dot{y} = 0$ and $\ddot{y} = 0$, the above expressions reduce to:

\begin{equation}\label{eq5.7}
\sigma_0 = -\frac{1}{4\pi y_0} \left[ 
\sqrt{1 - \frac{2M}{y_0} + \frac{Q^2}{y_0^2}} 
- \sqrt{1 - \frac{b(y_0)}{y_0}} 
\right],
\end{equation}

\begin{equation}\label{eq5.8}
P_0 = \frac{1}{8\pi y_0} \left[
\frac{1 - \frac{M}{y_0}}{\sqrt{1 - \frac{2M}{y_0} + \frac{Q^2}{y_0^2}}}
- \frac{\{1 + y_0 \Phi'(y_0)\}\left(1 - \frac{b(y_0)}{y_0}\right)}{\sqrt{1 - \frac{b(y_0)}{y_0}}}
\right].
\end{equation}

These relations explicitly determine the matter content localized at the junction surface. We shall use these two equations to study the following four important energy conditions \cite{Paramanik_2024,PARAMANIK_2025}:

\begin{itemize}
    \item \textbf{NEC}: \( \sigma_0 + P_0 \geq 0 \),
    \item \textbf{WEC}: \( \sigma_0 \geq 0 \quad \text{and} \quad \sigma_0 + P_0 \geq 0 \),
    \item \textbf{DEC}: \( \sigma_0 \geq 0 \quad \text{and} \quad \sigma_0 - |P_0| \geq 0 \),
    \item \textbf{SEC}: \( \sigma_0 + P_0 \geq 0 \quad \text{and} \quad \sigma_0 + 2 P_0 \geq 0 \).
\end{itemize}

In most wormhole models, the surface energy density is negative, indicating a violation of the null energy condition (NEC), which is a well-known and generic feature of traversable wormholes \cite{metric, Visser_1995}. Violation of NEC implies the presence of exotic matters in the spacetime, whereas violation of SEC leads to a gravity which is repulsive in nature. The presence of an electric charge introduces an additional contribution to the effective gravitational field, which can significantly affect the configuration's energy requirements and stability. In particular, the repulsive nature of the electromagnetic field may partially offset gravitational attraction, thereby reducing the extent of exotic matter required at the shell.


\section{Solutions of the field equations}\label{sec6}

In this section, we shall be trying to solve \cref{eq4.8}, and while doing so, we notice that this equation contains two unknown functions, namely $b(r)$ and $\Phi(r)$. Thus, our strategy is to treat any well-known function as one of the unknowns and then derive the other unknown function by solving the resulting equation. We shall be choosing three such well documented forms of the redshift function $\Phi(r)$. After obtaining a solution, we shall conduct a thorough graphical analysis of the resulting model to study the properties of wormholes and the validity of the energy conditions associated with them. For the remainder of the text, we have chosen $r_0 = 1,~ M = 0.5,~ \text{and} ~ Q = 0.3$, which gives us the outer horizon radius $r_+ = 0.9$ and so we get $r_+ < r_0$ which is the desired setup for a thin--shell wormhole. We also have studied three different epochs of cosmic time by invoking $z=0.5,~ z=0, ~\text{and}~ z=-0.5$ corresponding to the past, present and future respectively in our obtained solutions. While performing graphical analysis, we have chosen the best-fit values for different parameters from \cref{tab:bestfit_parameters}.

\subsection{Model I: $\Phi(r) = \text{constant}$}

We begin by considering the redshift function as constant. Thus, when $\Phi(r) = $ constant, we get from solving \cref{eq4.8} while also considering $b(r_0) = r_0$:

\begin{equation}\label{eq5.1.1}
\begin{aligned}
    b(r) ~= -\frac{3 \alpha  \gamma  r^3-r_0 \left(\frac{r}{r_0}\right)^{-\frac{\sqrt{\alpha ^2-4 \gamma }}{\alpha }}  \left\{8 \pi  (\beta -1) \left(\sqrt{\alpha ^2-4 \gamma }+3 \alpha \right)+3 \alpha  \gamma  r_0^2\right\}}{8 \pi  (\beta -1) \left(\sqrt{\alpha ^2-4 \gamma }+3 \alpha \right)}
\end{aligned}
\end{equation}

\begin{figure}[htbp]
\centering
\begin{subfigure}{0.49\textwidth}
    \centering
    \includegraphics[width=0.98\textwidth]{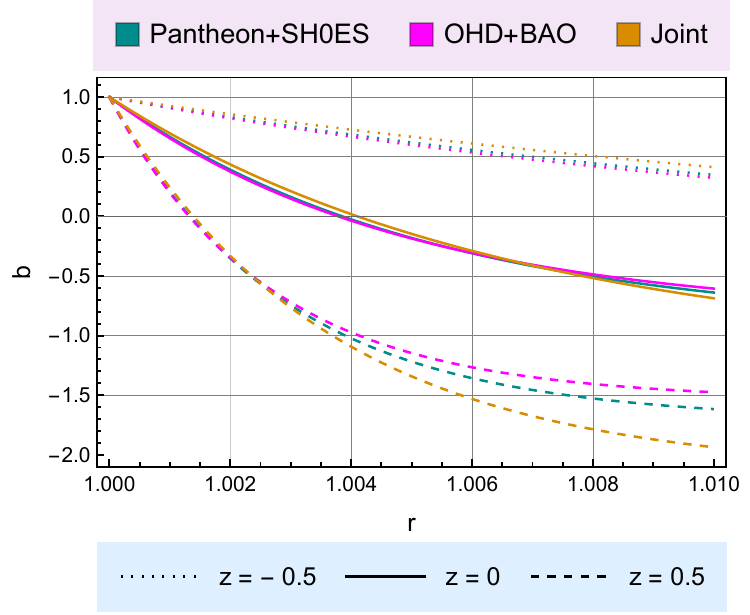}
    \caption{}
\end{subfigure}
\hfill
\begin{subfigure}{0.49\textwidth}
    \centering
    \includegraphics[width=0.98\textwidth]{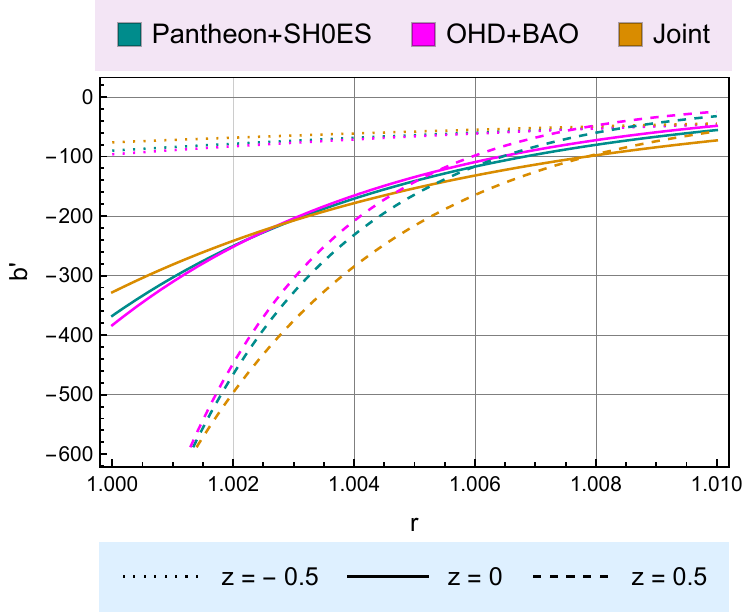}
    \caption{}
\end{subfigure}
\par\bigskip
\begin{subfigure}{0.49\textwidth}
    \centering
    \includegraphics[width=0.98\textwidth]{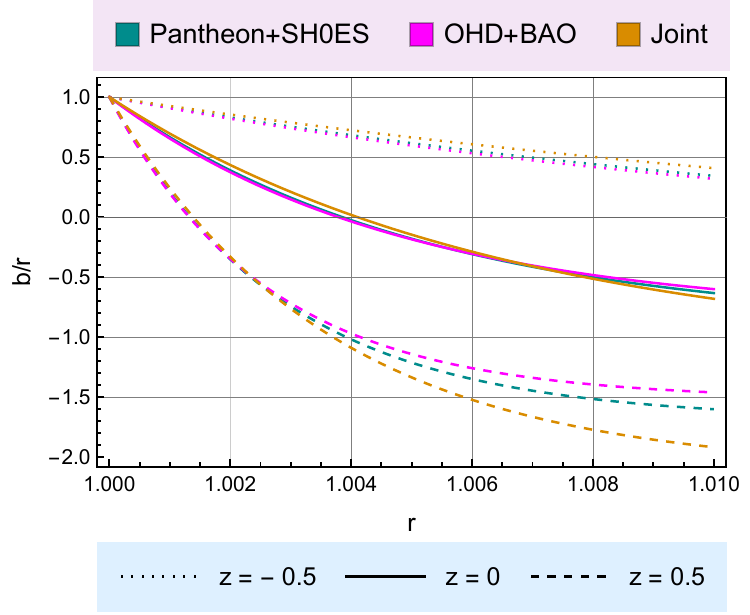}
    \caption{}
\end{subfigure}
\hfill
\begin{subfigure}{0.49\textwidth}
    \centering
    \includegraphics[width=0.98\textwidth]{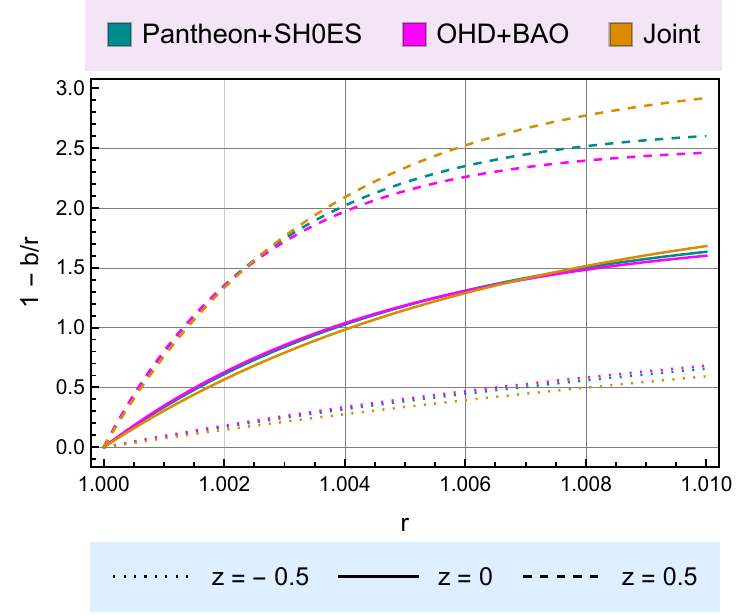}
    \caption{}
\end{subfigure}
\par\bigskip
\begin{subfigure}{0.98\textwidth}
    \centering
    \includegraphics[width=0.49\textwidth]{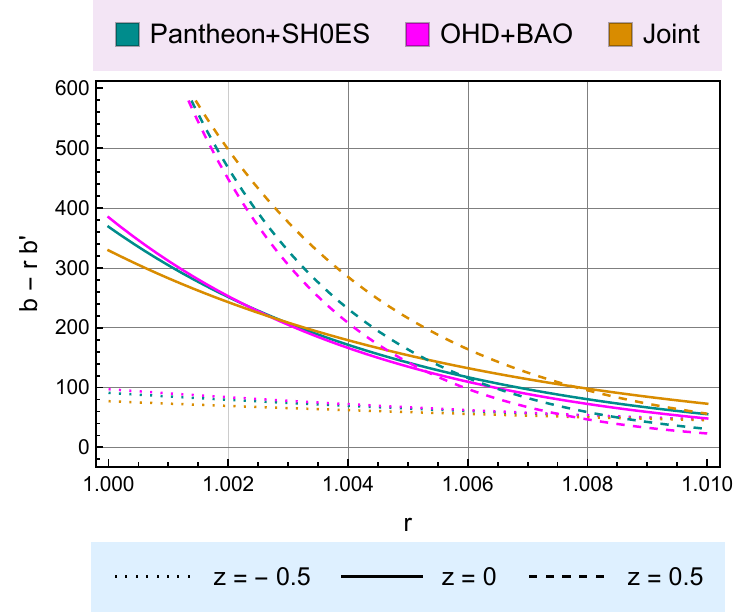}
    \caption{}
\end{subfigure}
\caption{Shape function and various properties of it for \textit{Model I} when $M = 0.5,~ Q = 0.3,~ r_0 = 1$}
\label{fig3}
\end{figure}

\begin{figure}[htbp]
\centering
\begin{subfigure}{0.49\textwidth}
    \centering
    \includegraphics[width=0.98\textwidth]{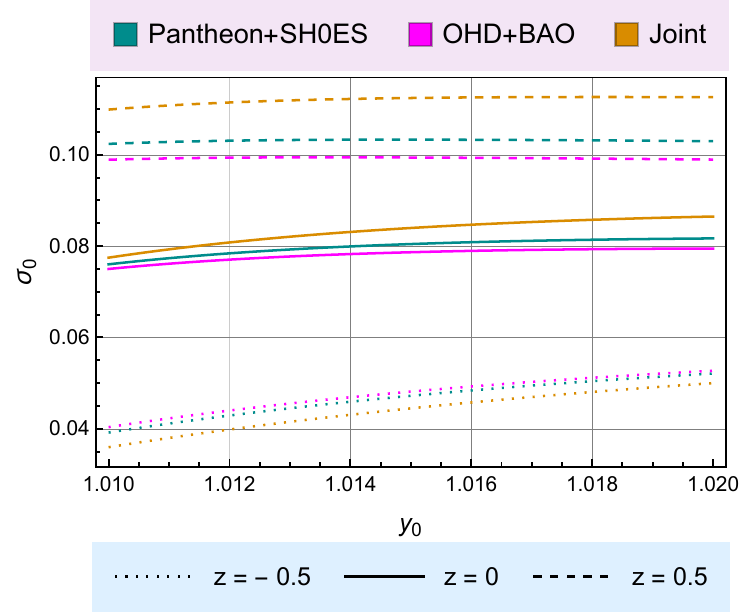}
    \caption{}
\end{subfigure}
\hfill
\begin{subfigure}{0.49\textwidth}
    \centering
    \includegraphics[width=0.98\textwidth]{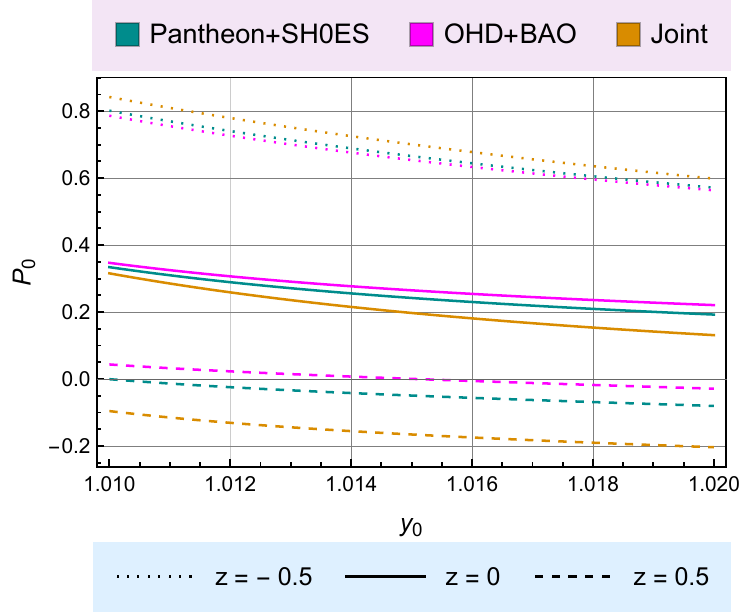}
    \caption{}
\end{subfigure}
\par\bigskip
\begin{subfigure}{0.49\textwidth}
    \centering
    \includegraphics[width=0.98\textwidth]{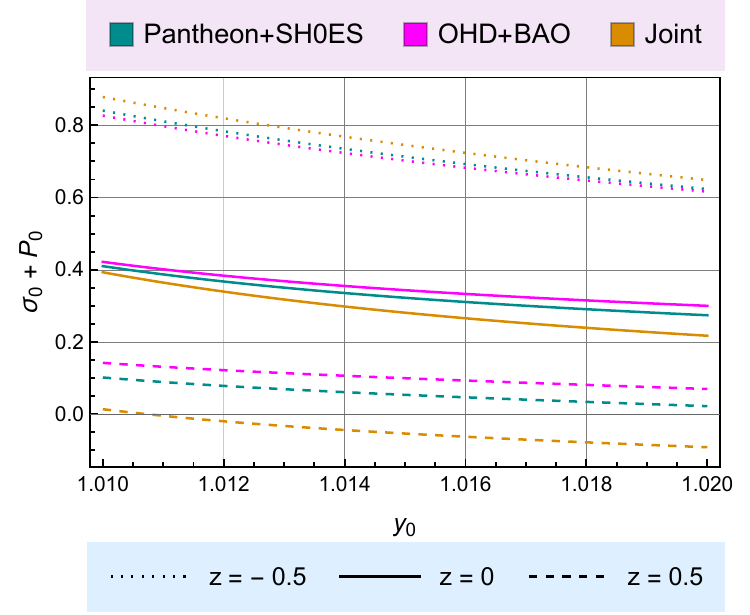}
    \caption{}
\end{subfigure}
\hfill
\begin{subfigure}{0.49\textwidth}
    \centering
    \includegraphics[width=0.98\textwidth]{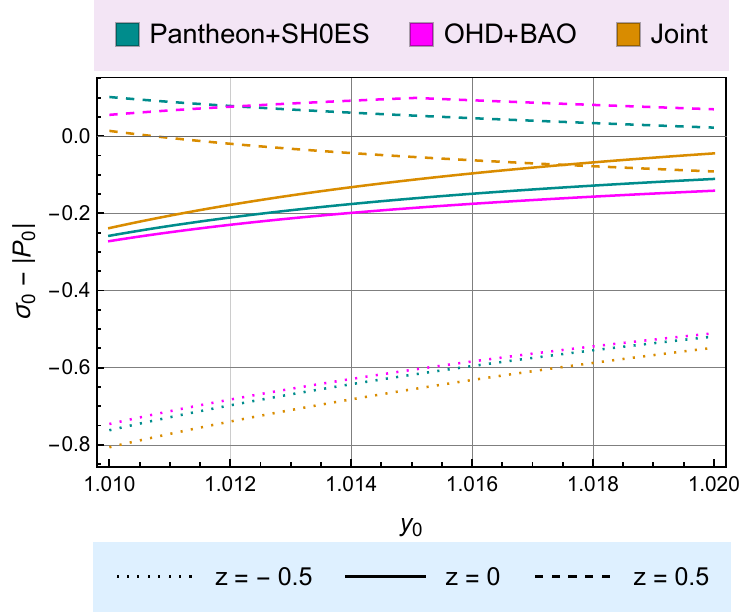}
    \caption{}
\end{subfigure}
\par\bigskip
\begin{subfigure}{0.98\textwidth}
    \centering
    \includegraphics[width=0.49\textwidth]{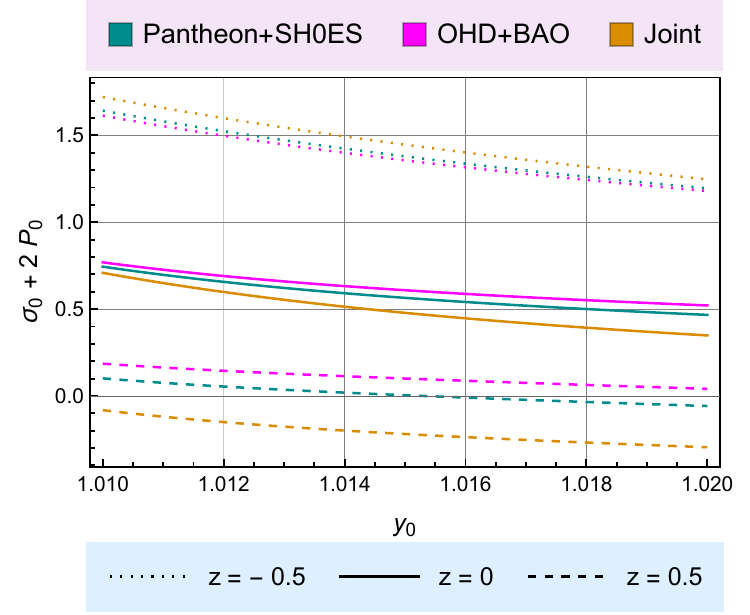}
    \caption{}
\end{subfigure}
\caption{Energy conditions for \textit{Model I} when $M = 0.5,~ Q = 0.3$}
\label{fig4}
\end{figure}

A graphical analysis of this solution for the shape function is shown in \cref{fig3}. We immediately notice that for $z=0$ and $z=0.5$, none of the traversability conditions are satisfied by this solution. So the resulting wormhole neither flares out nor is it asymptotically flat. Hence, we deploy the cut-paste method described in the previous section and construct a thin--shell wormhole by choosing the junction radius $y_0 \geq 1.01$. We then have graphically studied the associated energy conditions in \cref{fig4}. We see that the surface energy $\sigma$ is positive for all three values of $z$ and all three sets of best-fit values. The surface pressure $P$, on the other hand, is positive for $z=-0.5$ and $z=0$, and is mostly negative for $z=0.5$. The graphs for $\sigma+P$ and $\sigma+2P$ show outcomes similar to those of the surface pressure graphs, while the $\sigma-|P|$ graphs are different in the sense that they are negative for $z=-0.5$ and $z=0$ and are mostly positive for $z=0.5$. From all these observations, we can conclude that the NEC, WEC, and SEC are satisfied and DEC is violated for $z=-0.5$ and $z=0$, whereas all these energy conditions can be violated for some domain of $y$ when $z=0.5$.


\subsection{Model II: $\displaystyle \Phi(r)= - \frac{r_0}{r}$}

In \textit{Model II}, we consider $\Phi(r) = - (r_0/r)$ as the redshift function and thus solving \cref{eq4.8} with the same initial condition yields:

\begin{equation}
\begin{aligned}
    b(r) ~&= \frac{\exp\left(-\frac{\sqrt{\alpha ^2-4 \gamma } \left\{\log (r)-\frac{2 r_0}{r}\right\}}{\alpha}\right)}{\alpha ^2 \pi (\beta - 1)} \left[ \pi  \alpha ^2 (\beta -1) r_0 \exp\left({\frac{\sqrt{\alpha ^2-4 \gamma } \left\{\log \left(r_0\right)-2\right\}}{\alpha }}\right)\right. \\
    & - 2^{\frac{\sqrt{\alpha ^2-4 \gamma }}{\alpha }} \left(\frac{\sqrt{\alpha ^2-4 \gamma }}{\alpha }\right)^{\frac{\sqrt{\alpha ^2-4 \gamma }+\alpha }{\alpha }} r_0^{\frac{\sqrt{\alpha ^2-4 \gamma }+\alpha }{\alpha}} \left\{ 2 \pi  \alpha ^2 (\beta -1) \Gamma \left(-\frac{\sqrt{\alpha ^2-4 \gamma }}{\alpha },\frac{2 \sqrt{\alpha ^2-4 \gamma }}{\alpha }\right)\right. \\
    & \left. -3 \gamma  r_0^2 \left(\alpha ^2-4 \gamma \right) \Gamma \left(-\frac{\sqrt{\alpha ^2-4 \gamma }}{\alpha }-3,\frac{2 \sqrt{\alpha ^2-4 \gamma }}{\alpha }\right)\right\} - 2^{\frac{\sqrt{\alpha ^2-4 \gamma }}{\alpha }} r^{\frac{\sqrt{\alpha ^2-4 \gamma }+\alpha }{\alpha }} \left(\frac{r_0 \sqrt{\alpha ^2-4 \gamma }}{\alpha  r}\right){}^{\frac{\sqrt{\alpha ^2-4 \gamma }+\alpha }{\alpha }} \\
    & \times \left.\left\{ 3 \gamma  r_0^2 \left(\alpha ^2-4 \gamma \right) \Gamma \left(-\frac{\sqrt{\alpha ^2-4 \gamma }}{\alpha }-3,\frac{2 \sqrt{\alpha ^2-4 \gamma } r_0}{r \alpha }\right)-2 \pi  \alpha ^2 (\beta -1) \Gamma \left(-\frac{\sqrt{\alpha ^2-4 \gamma }}{\alpha },\frac{2 \sqrt{\alpha ^2-4 \gamma } r_0}{r \alpha }\right) \right\} \right] ~.
\end{aligned}
\end{equation}

\begin{figure}[htbp]
\centering
\begin{subfigure}{0.49\textwidth}
    \centering
    \includegraphics[width=0.98\textwidth]{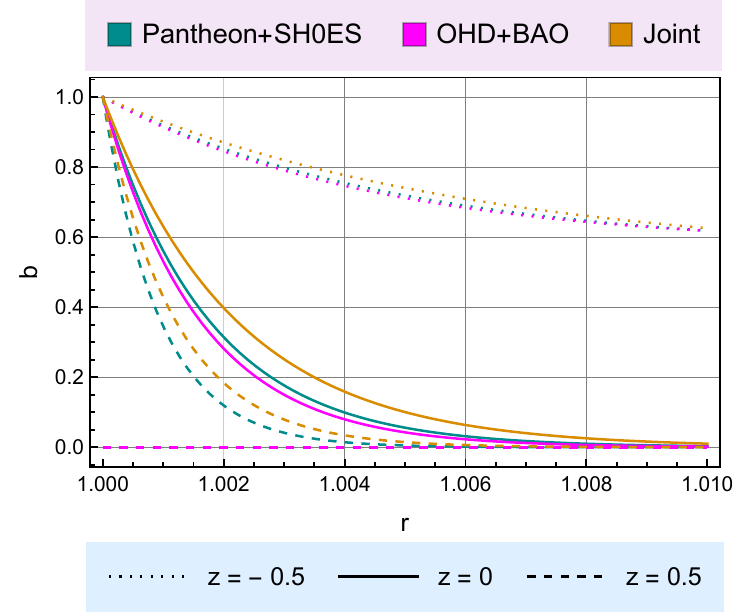}
    \caption{}
\end{subfigure}
\hfill
\begin{subfigure}{0.49\textwidth}
    \centering
    \includegraphics[width=0.98\textwidth]{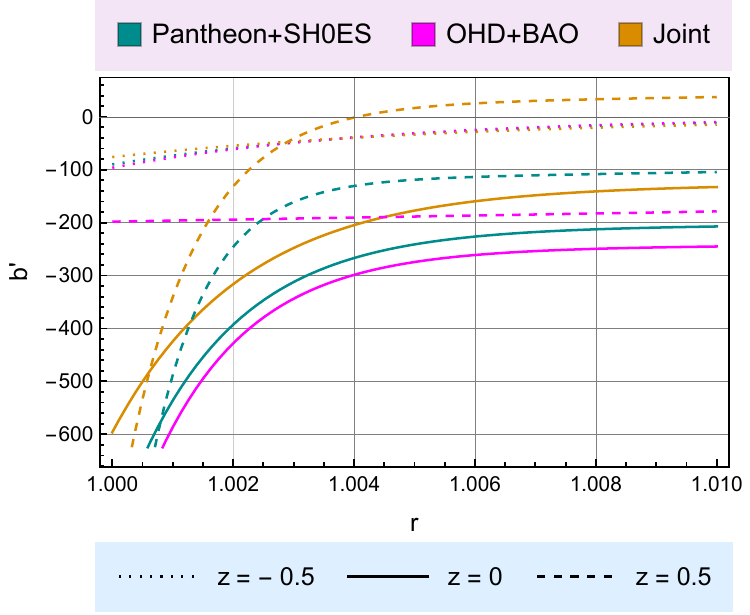}
    \caption{}
\end{subfigure}
\par\bigskip
\begin{subfigure}{0.49\textwidth}
    \centering
    \includegraphics[width=0.98\textwidth]{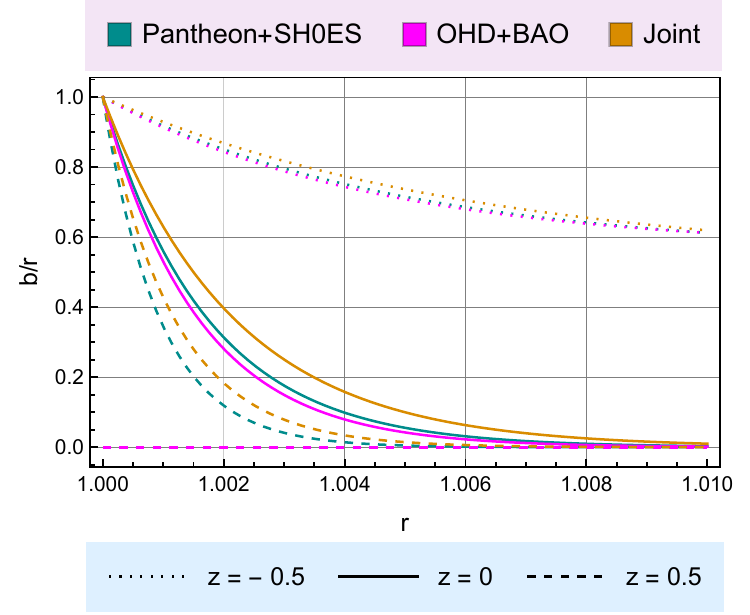}
    \caption{}
\end{subfigure}
\hfill
\begin{subfigure}{0.49\textwidth}
    \centering
    \includegraphics[width=0.98\textwidth]{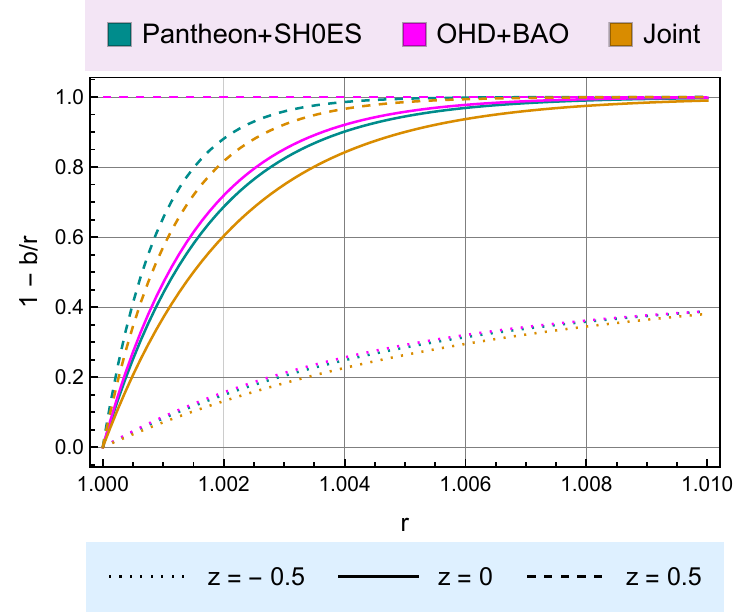}
    \caption{}
\end{subfigure}
\par\bigskip
\begin{subfigure}{0.98\textwidth}
    \centering
    \includegraphics[width=0.49\textwidth]{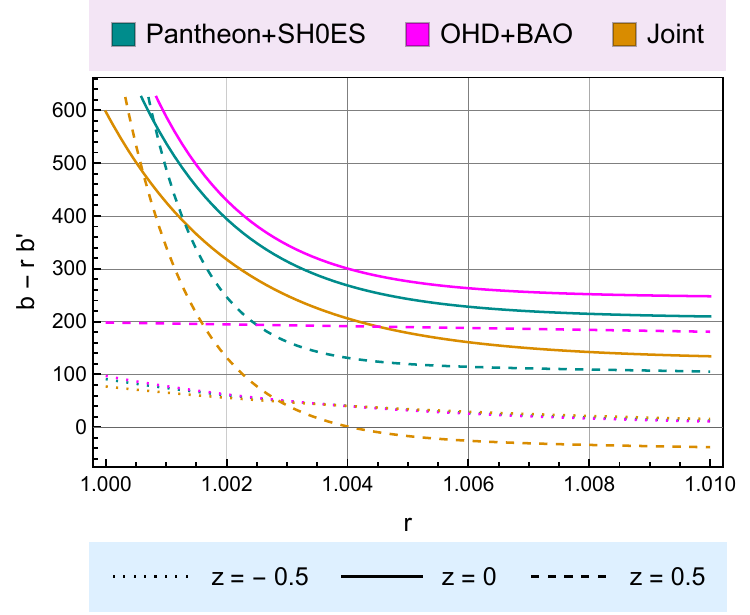}
    \caption{}
\end{subfigure}
\caption{Shape function and various properties of it for \textit{Model II} when $M = 0.5,~ Q = 0.3,~ r_0 = 1$}
\label{fig5}
\end{figure}

\begin{figure}[htbp]
\centering
\begin{subfigure}{0.49\textwidth}
    \centering
    \includegraphics[width=0.98\textwidth]{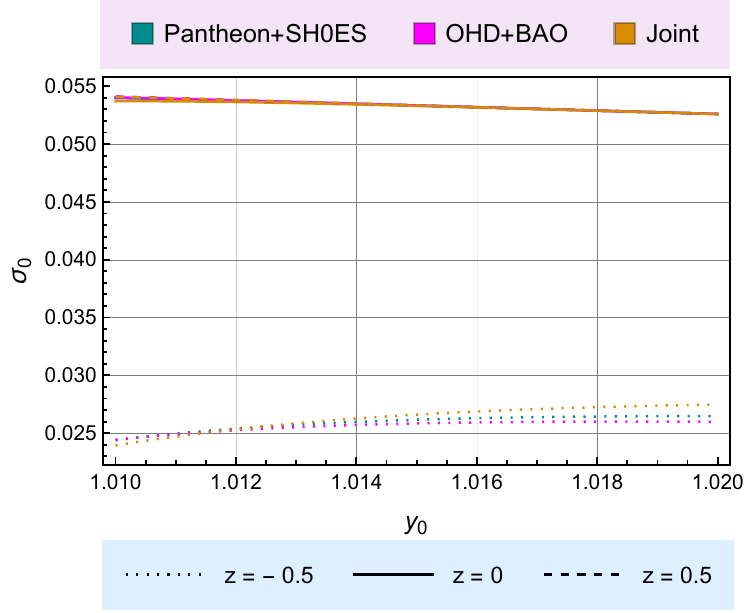}
    \caption{}
\end{subfigure}
\hfill
\begin{subfigure}{0.49\textwidth}
    \centering
    \includegraphics[width=0.98\textwidth]{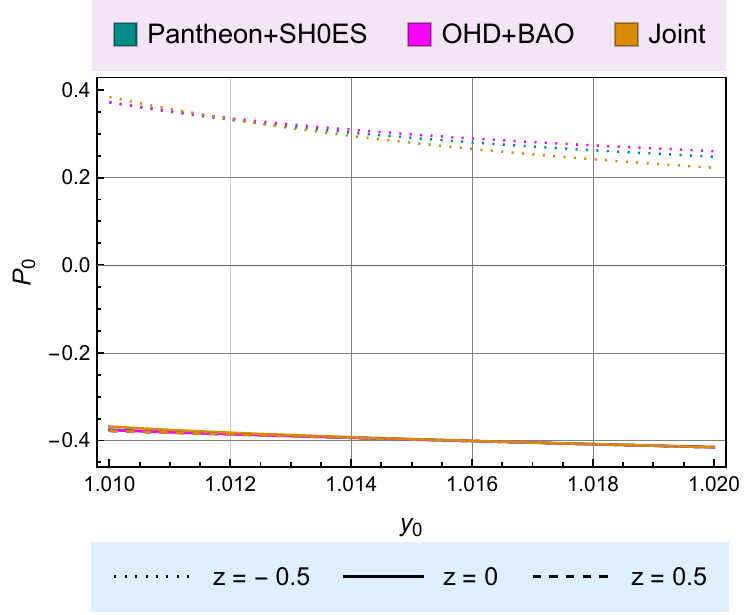}
    \caption{}
\end{subfigure}
\par\bigskip
\begin{subfigure}{0.49\textwidth}
    \centering
    \includegraphics[width=0.98\textwidth]{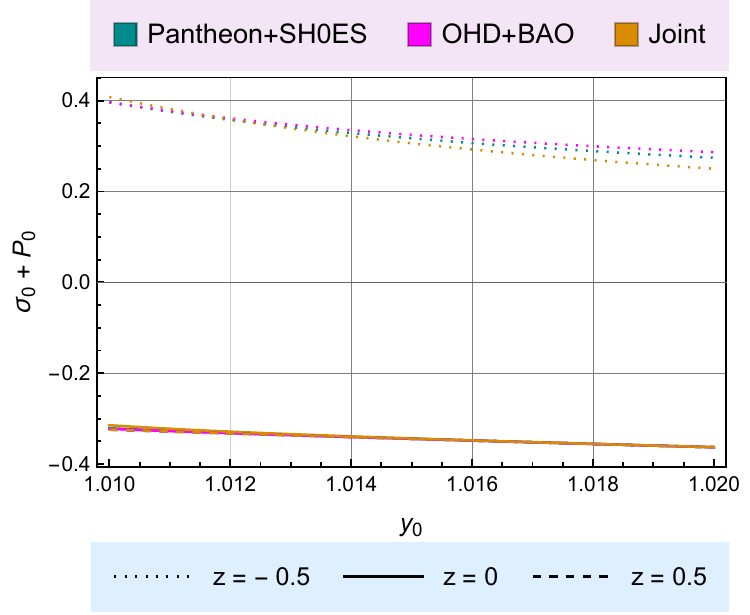}
    \caption{}
\end{subfigure}
\hfill
\begin{subfigure}{0.49\textwidth}
    \centering
    \includegraphics[width=0.98\textwidth]{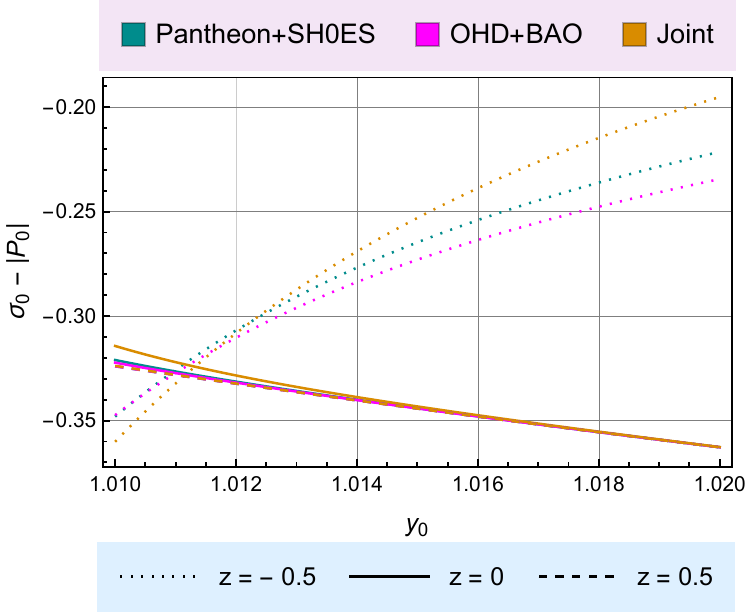}
    \caption{}
\end{subfigure}
\par\bigskip
\begin{subfigure}{0.98\textwidth}
    \centering
    \includegraphics[width=0.49\textwidth]{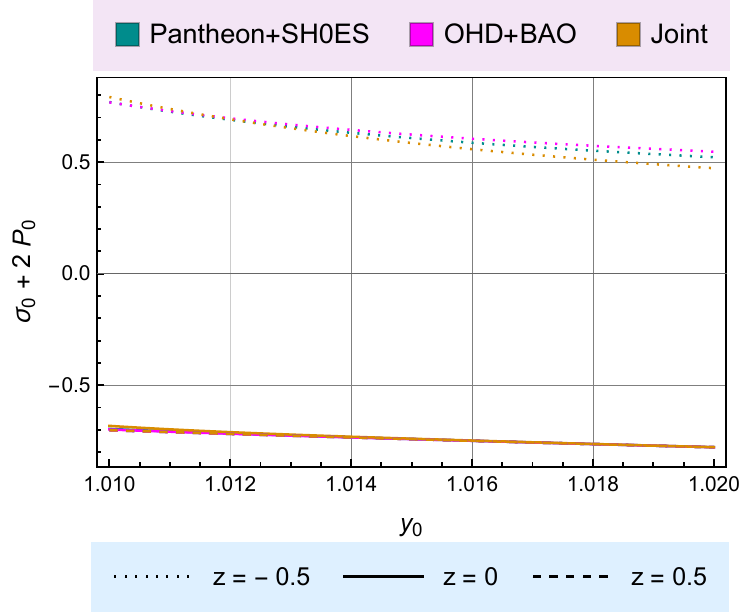}
    \caption{}
\end{subfigure}
\caption{Energy conditions for \textit{Model II} when $M = 0.5,~ Q = 0.3$}
\label{fig6}
\end{figure}

As before, we have performed a similar graphical analysis for this model, which is shown in \cref{fig5}. We notice that all the traversability conditions are met by this model when $ z=0$ or $0.5$, but it breaks down when $z=-0.5$. Using the cut--paste method, we have constructed a thin--shell wormhole structure and studied the corresponding energy conditions graphically in \cref{fig6}. From \cref{fig6}, we can easily conclude that this thin--shell structure breaks down all four energy conditions when $z=0$ and $z=0.5$, and only NEC, WEC, and SEC are satisfied for the $z=-0.5$ case.


\subsection{Model III: $\displaystyle \Phi(r) = \log \left(\frac{r}{r_0}\right)$}

Finally, we choose $\Phi(r) = \log (r/r_0)$ for \textit{Model III}, and so we obtain from \cref{eq4.8} while also considering $b(r_0) = r_0$ the following solution:

\begin{equation}
\begin{aligned}
    b(r) ~&= \frac{1}{32 \pi  (\beta -1) \left\{\alpha  \left(\sqrt{\alpha ^2-4 \gamma }+\alpha \right)-3 \gamma \right\}}\left[ 16 \pi r (\beta -1) \left\{\alpha  \left(\sqrt{\alpha ^2-4 \gamma }+\alpha \right)-4 \gamma \right\}-\alpha  \gamma  r^3 \left(3 \sqrt{\alpha ^2-4 \gamma }+\alpha \right)\right. \\
    & \left. +r^{-\frac{3 \sqrt{\alpha ^2-4 \gamma }}{\alpha }} r_0^{\frac{3 \sqrt{\alpha ^2-4 \gamma }}{\alpha }+1} \left\{16 \pi  (\beta -1) \left(\alpha  \left(\sqrt{\alpha ^2-4 \gamma }+\alpha \right)-2 \gamma \right)+\alpha  \gamma  r_0^2 \left(3 \sqrt{\alpha ^2-4 \gamma }+\alpha \right)\right\} \right] ~.
\end{aligned}
\end{equation}

\begin{figure}[htbp]
\centering
\begin{subfigure}{0.49\textwidth}
    \centering
    \includegraphics[width=0.98\textwidth]{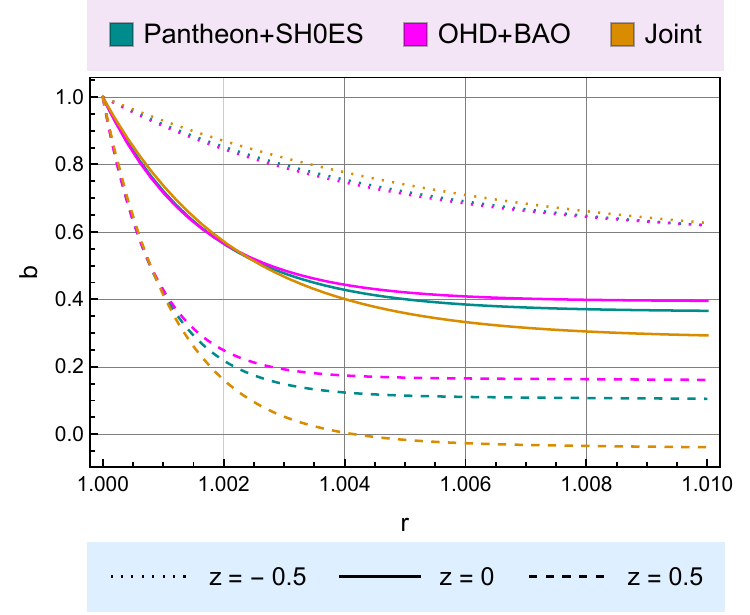}
    \caption{}
\end{subfigure}
\hfill
\begin{subfigure}{0.49\textwidth}
    \centering
    \includegraphics[width=0.98\textwidth]{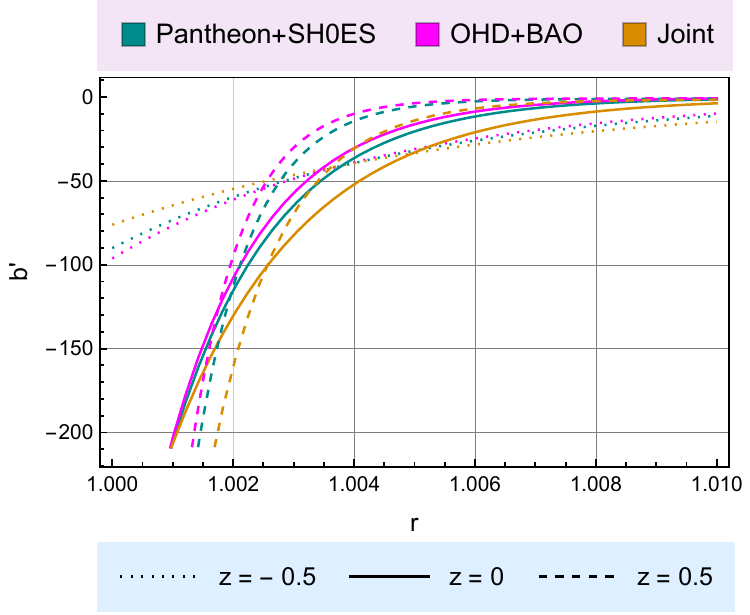}
    \caption{}
\end{subfigure}
\par\bigskip
\begin{subfigure}{0.49\textwidth}
    \centering
    \includegraphics[width=0.98\textwidth]{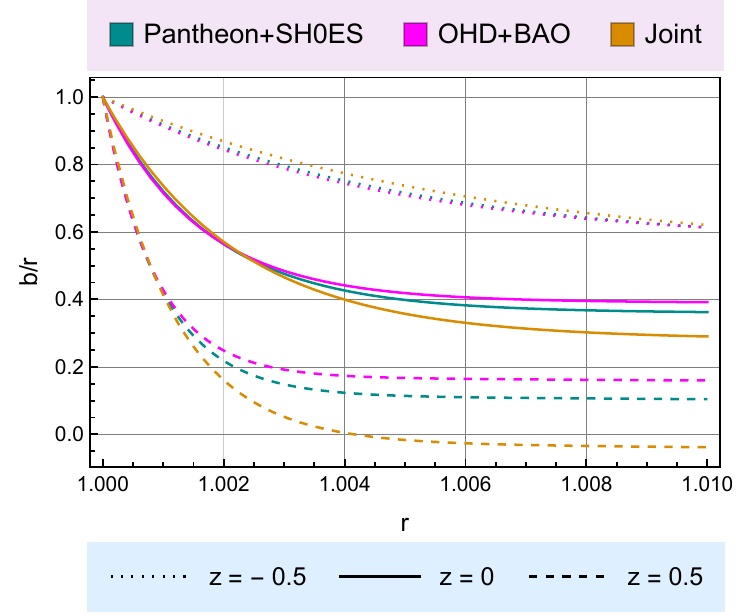}
    \caption{}
\end{subfigure}
\hfill
\begin{subfigure}{0.49\textwidth}
    \centering
    \includegraphics[width=0.98\textwidth]{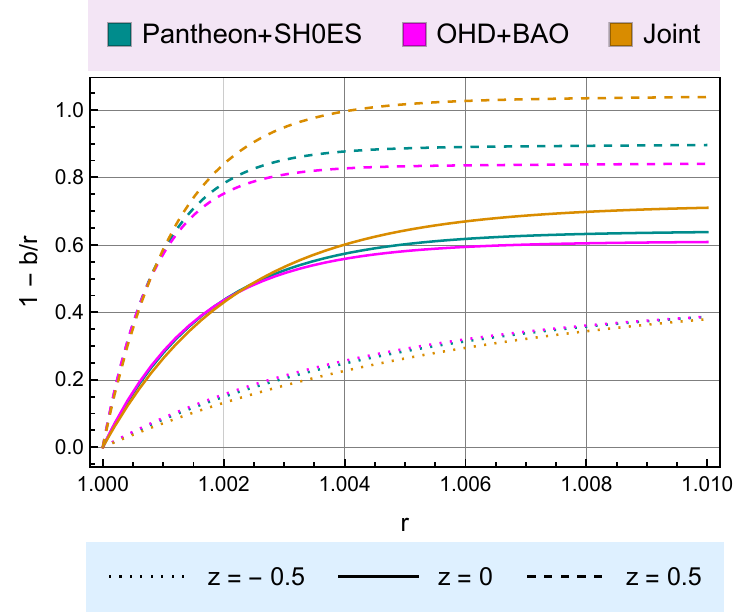}
    \caption{}
\end{subfigure}
\par\bigskip
\begin{subfigure}{0.98\textwidth}
    \centering
    \includegraphics[width=0.49\textwidth]{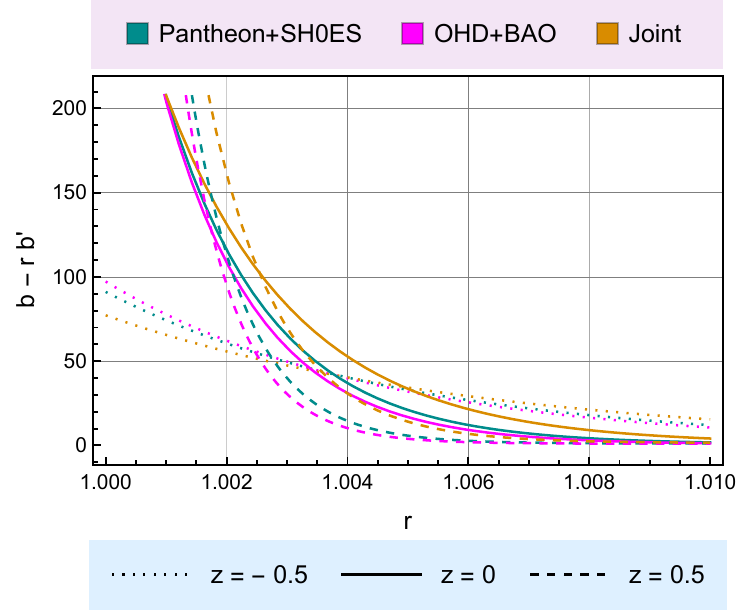}
    \caption{}
\end{subfigure}
\caption{Shape function and various properties of it for \textit{Model III} when $M = 0.5,~ Q = 0.3,~ r_0 = 1$}
\label{fig7}
\end{figure}

\begin{figure}[htbp]
\centering
\begin{subfigure}{0.49\textwidth}
    \centering
    \includegraphics[width=0.98\textwidth]{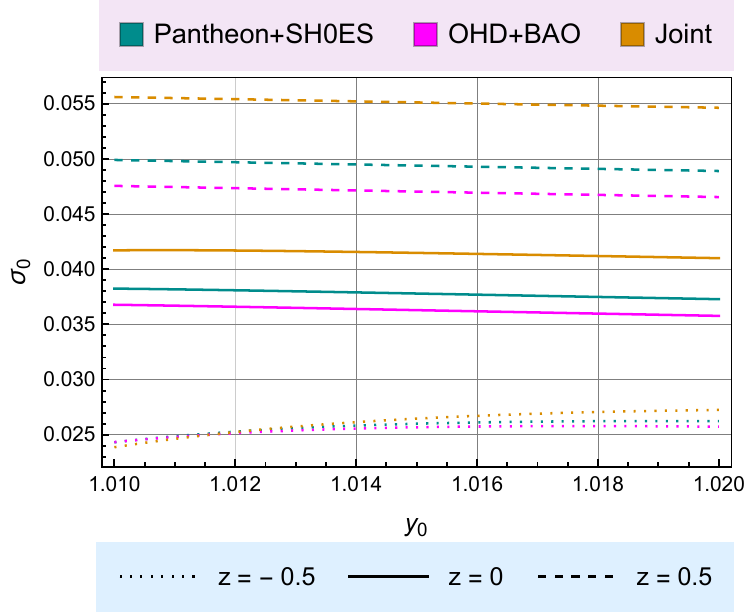}
    \caption{}
\end{subfigure}
\hfill
\begin{subfigure}{0.49\textwidth}
    \centering
    \includegraphics[width=0.98\textwidth]{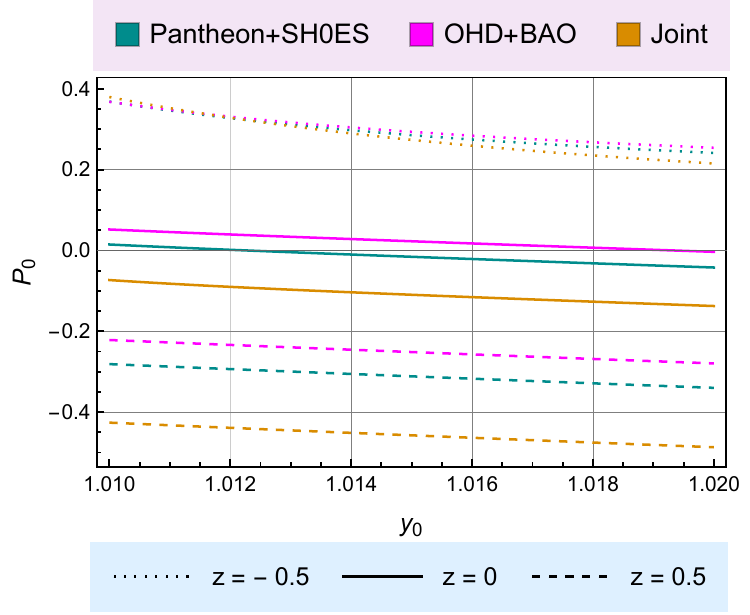}
    \caption{}
\end{subfigure}
\par\bigskip
\begin{subfigure}{0.49\textwidth}
    \centering
    \includegraphics[width=0.98\textwidth]{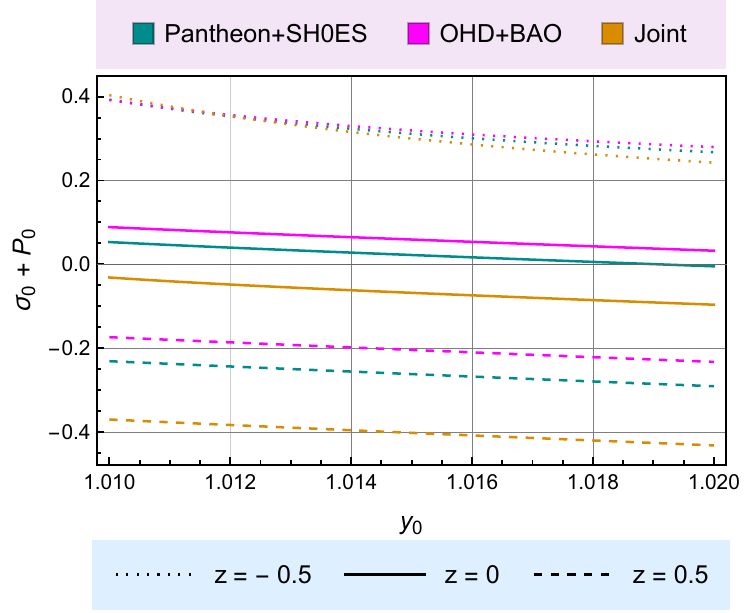}
    \caption{}
\end{subfigure}
\hfill
\begin{subfigure}{0.49\textwidth}
    \centering
    \includegraphics[width=0.98\textwidth]{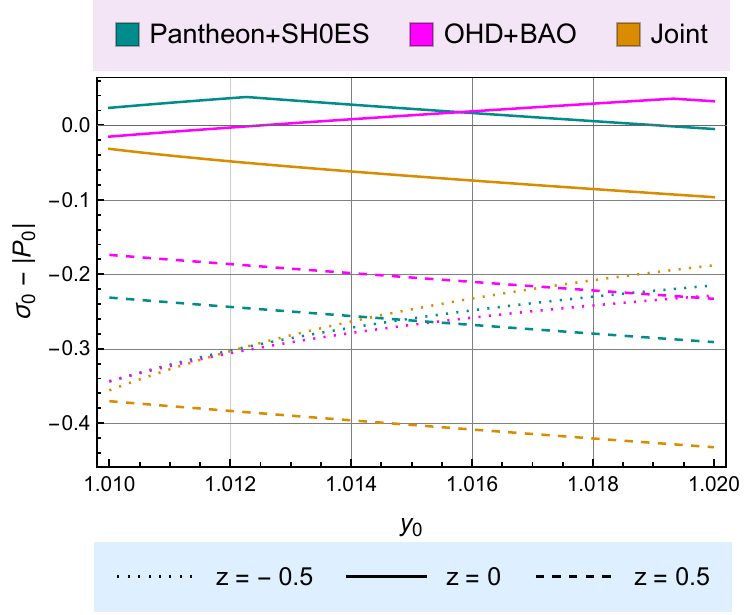}
    \caption{}
\end{subfigure}
\par\bigskip
\begin{subfigure}{0.98\textwidth}
    \centering
    \includegraphics[width=0.49\textwidth]{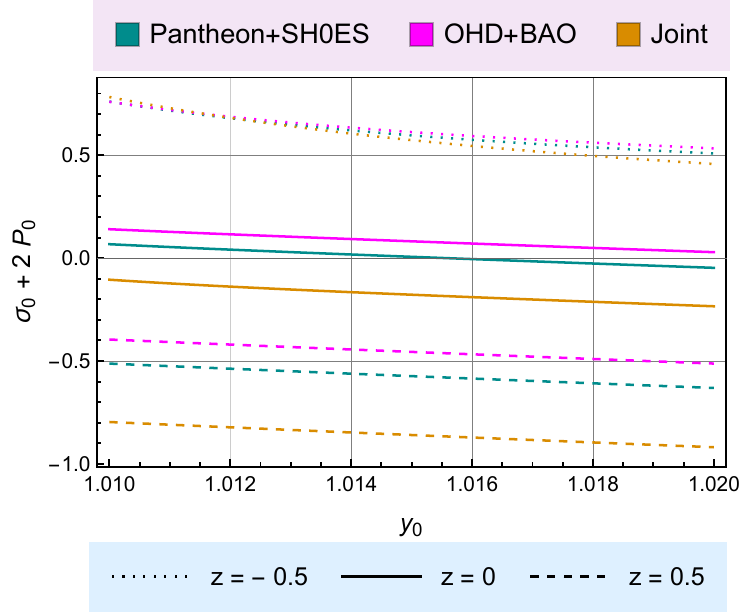}
    \caption{}
\end{subfigure}
\caption{Energy conditions for \textit{Model III} when $M = 0.5,~ Q = 0.3$}
\label{fig8}
\end{figure}

When plotted in \cref{fig7}, we immediately notice that this shape function is not asymptotically flat for any configuration. So, we again match the interior wormhole spacetime to an exterior asymptotically flat spacetime, as in the previous two models. Graphical analysis of the associated energy conditions in \cref{fig8} also reveals that NEC, WEC, and SEC are satisfied by the thin--shell structure when $z=-0.5$ and violated when $z=0.5$, and DEC is violated for these two cases. The $z=0$ case shows mixed outcomes across all four energy conditions, \textit{i.e.}, for some domain of $y$ they are satisfied, whereas they are violated in others.


\section{Linearized stability analysis of the thin--shell}\label{sec7}

The stability of the junction against radial perturbations can be investigated by allowing the matching radius to depend on the proper time, $y=y(\tau)$. The dynamics of the thin shell may then be conveniently described by an effective potential. For the present configuration, we introduce the metric functions:

\begin{equation}\label{eq6.1}
    f_{+}(y) = 1-\frac{2M}{y}+\frac{Q^{2}}{y^{2}} ~,
    \qquad
    f_{-}(y) = 1-\frac{b(y)}{y} ~,
\end{equation}
which correspond, respectively, to the exterior Reissner--Nordstr{\"o}m geometry and the interior Morris--Thorne spacetime. The dynamical surface energy density at the junction is given by modifying \cref{eq5.5} as follows:

\begin{equation}\label{eq6.2}
    \sigma(y) = -\frac{1}{4\pi y} \left[ \sqrt{f_{+}(y)+\dot{y}^{\,2}} - \sqrt{f_{-}(y)+\dot{y}^{\,2}} \right] ~.
\end{equation}
It is useful to introduce the surface mass of the junction:

\begin{equation}\label{eq6.3}
    m_s(y)=4\pi y^{2}\sigma(y) ~,
\end{equation}
together with the auxiliary functions:

\begin{equation}\label{eq6.4}
    F(y)=\frac{f_{+}(y)+f_{-}(y)}{2} ~,
    \qquad
    G(y)=\frac{f_{+}(y)-f_{-}(y)}{2} ~.
\end{equation}
The radial equation of motion can consequently be cast into the form \cite{Manna2026tsw, Sarkar2026tsw}:

\begin{equation}\label{eq6.5}
    \dot{y}^{\,2}+V(y)=0 ~,
\end{equation}
where the effective potential is:

\begin{equation}\label{eq6.6}
    V(y) = F(y) -\frac{m_s^{2}(y)}{4y^{2}} -\frac{y^{2}G^{2}(y)}{m_s^{2}(y)} ~.
\end{equation}
For the geometries considered here, the functions $F(y)$ and $G(y)$ take the following explicit forms:

\begin{equation}\label{eq6.7}
    F(y) = 1-\frac{M}{y} +\frac{Q^{2}}{2y^{2}} -\frac{b(y)}{2y} ~,
    \qquad
    G(y) = \frac{b(y)-2M}{2y} +\frac{Q^{2}}{2y^{2}} ~.
\end{equation}

Thus, once the background geometry and the response of the surface matter are specified, the radial dynamics of the junction is entirely encoded in the effective potential $V(y)$.

For a transparent shell, the momentum-flux term across the junction vanishes, $\Xi=0$, and the conservation identity reduces to:

\begin{equation}\label{eq6.8}
    \sigma'(y) = -\frac{2}{y} \left[ \sigma(y)+P(y) \right] ~,
\end{equation}
where a prime denotes differentiation with respect to $y$. We now consider a static configuration in which the surface energy density and pressure are given by \cref{eq5.7} and \cref{eq5.8}, respectively. We also incorporate \cref{eq4.7} as the \textit{GGDE} equation of state to solve \cref{eq6.8}. Naturally, to utilize the \cref{eq4.7}, we have to replace $\rho$ with $\sigma$ for surface density and $p$ with $P$ for pressure. The solution of \cref{eq6.8} in the context of \textit{GGDE} now reads:

\begin{equation}\label{eq6.9}
    \sigma(y) = \frac{y^{-\frac{2 \alpha }{\sqrt{\alpha ^2-4 \gamma }}-2} \left[y_0^{\frac{2 \alpha }{\sqrt{\alpha ^2-4 \gamma }}+2} \left\{8 \pi \sigma _0 (\beta -1) \left(\sqrt{\alpha ^2-4 \gamma }+\alpha \right)+3 \alpha  \gamma \right\}-3 \alpha  \gamma  y^{\frac{2 \alpha }{\sqrt{\alpha ^2-4 \gamma }}+2}\right]}{8 \pi  (\beta -1) \left(\sqrt{\alpha ^2-4 \gamma }+\alpha \right)} ~,
\end{equation}
where $\sigma_0$ is expressed in \cref{eq5.7}. This relation will be used in \cref{eq6.3} for further calculations.

To determine the stability of the static configuration, we expand the effective potential about $y=y_{0}$ up to second order:

\begin{equation}\label{eq6.10}
    V(y) = V(y_{0}) + V'(y_{0})(y-y_{0}) + \frac{1}{2}V''(y_{0})(y-y_{0})^{2} + \mathcal{O}\left[(y-y_{0})^{3}\right] ~.
\end{equation}
At the equilibrium configuration:

\begin{equation}\label{eq6.11}
    V(y_{0})=0 ~,
    \qquad
    V'(y_{0})=0 ~,
\end{equation}
so that \cref{eq6.10} becomes:

\begin{equation}\label{eq6.12}
    V(y) = \frac{1}{2}V''(y_{0})(y-y_{0})^{2} + \mathcal{O}\left[(y-y_{0})^{3}\right] ~.
\end{equation}
Accordingly, the radial equation of motion takes the form:

\begin{equation}\label{eq6.13}
    \dot{y}^{\,2} = -\frac{1}{2}V''(y_{0})(y-y_{0})^{2} + \mathcal{O}\left[(y-y_{0})^{3}\right] ~.
\end{equation}

The stability of the equilibrium configuration is therefore determined by the sign of $V''(y_{0})$. If $V''(y_{0})<0$, the effective potential possesses a local maximum at $y_{0}$, and an infinitesimal radial perturbation drives the junction away from its equilibrium position. Conversely, the configuration is linearly stable whenever $V(y)$ possesses a local minimum at $y_{0}$, \textit{i.e.},:

\begin{equation}\label{eq6.14}
    V''(y_{0})>0 ~.
\end{equation}

\begin{figure}[htbp]
\centering
\begin{subfigure}{0.49\textwidth}
    \centering
    \includegraphics[width=0.98\textwidth]{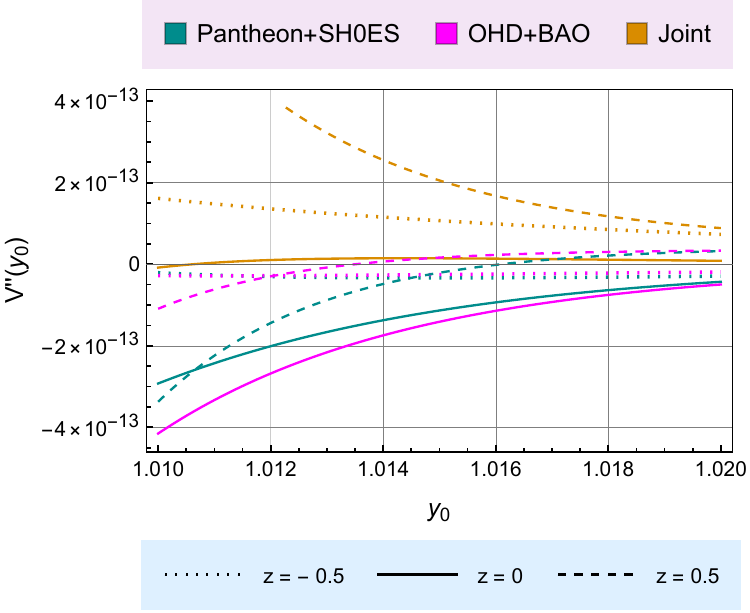}
    \caption{Model I}
\end{subfigure}
\hfill
\begin{subfigure}{0.49\textwidth}
    \centering
    \includegraphics[width=0.98\textwidth]{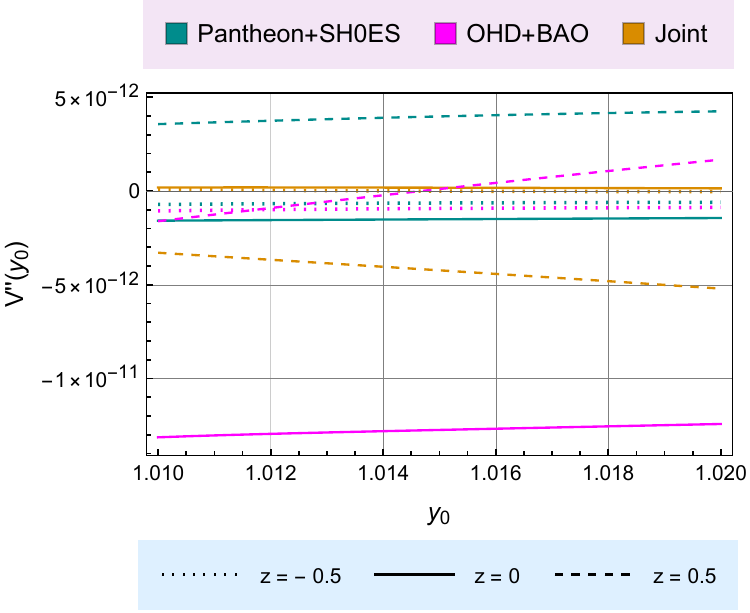}
    \caption{Model II}
\end{subfigure}
\par\bigskip
\begin{subfigure}{0.49\textwidth}
    \centering
    \includegraphics[width=0.98\textwidth]{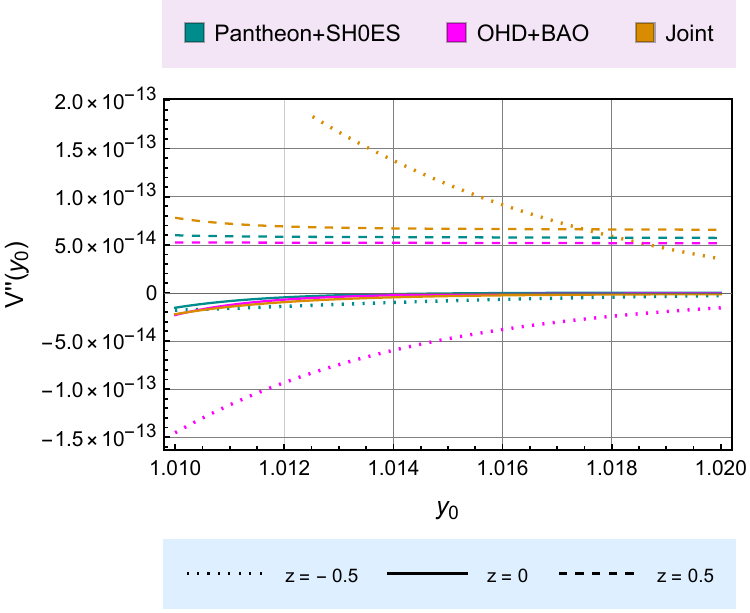}
    \caption{Model III}
\end{subfigure}
\caption{Stability analysis of thin--shell when $M = 0.5,~ Q = 0.3$}
\label{fig9}
\end{figure}

We have at first calculated $V''(y_0)$ using \cref{eq6.6} with the help of Eqs.~(\ref{eq6.3}), (\ref{eq6.7}), and (\ref{eq6.9}) and then have analyzed it's sign graphically in \cref{fig9}. Depending on the sign of $V''(y_0)$, we can conclude from the graphs that the thin--shells in \textit{Model I} and \textit{Model II} are mostly unstable, whereas in \textit{Model III}, all of the thin--shells corresponding to $z=0.5$ exhibit a stable nature. Nevertheless, it can also be seen that the thin--shells become more and more stable with increasing values of the junction radius $y_0$.


\section{Summary and conclusion}\label{sec8}

Throughout this work we have constructed a set of wormhole geometries supported exclusively by a GGDE source. The analysis opened with a concise summary of GGDE within a homogeneous and isotropic cosmological setting. We subsequently fixed the model's free parameters via a Markov Chain Monte Carlo (MCMC) fit to the Pantheon+SH0ES, OHD+BAO, and Joint datasets, all of which agree closely with one another. Attention then turned to the details of the wormhole geometry and the thin-shell formalism, from which several explicit wormhole configurations were obtained and their properties examined graphically. A brief stability analysis of the resulting thin--shell structures was carried out using the effective potential. The principal findings are summarized below:

\begin{enumerate}[(i)]
    \item The best-fit Hubble constant stays near $72\,\mathrm{km\,s^{-1}\,Mpc^{-1}}$, and the current matter density parameter comes out to $\Omega_{m0} \approx 0.27$ for every dataset considered. This mutual agreement among the estimates indicates that the model is insensitive to the particular choice of dataset and yields stable constraints on its parameters.

    \item As shown in Figs. \ref{fig:Hubble} and \ref{fig:distance modulus}, the model's theoretical predictions track the observational data closely, showing that it accounts for the measurements in a satisfactory manner.

    \item Fig. \ref{fig:getdist} shows the \texttt{GetDist} contour plots generated from the MCMC analysis for the CC+BAO, $Pantheon^+$, and combined datasets. They display the marginalized posterior distributions of the model parameters together with the associated confidence contours, bringing out the resulting parameter constraints and correlations for each dataset.

    \item Comparing the models via the AIC and BIC criteria shows that GGDE performs on par with the standard $\Lambda$CDM model, being favored under both the Pantheon+SH0ES and Joint datasets, whereas the OHD+BAO dataset gives only a marginal edge to $\Lambda$CDM.

    \item In addition, the Gelman-Rubin convergence diagnostic points to adequate convergence within the MCMC chains of the OHD+BAO and Joint analyses ($\hat{R}\approx1$), which confirms that the derived parameter constraints are robust and reliable.

    \item None of the three wormhole models fully meets every traversability requirement, as illustrated in Figs.~\ref{fig3}, \ref{fig5}, and \ref{fig7}. For \textit{Model I} and \textit{Model III}, the shape function does not meet the criteria for asymptotic flatness, while for \textit{Model II}, the $z=0.5$ configuration in the Joint dataset breaches the flare-out condition. To remedy this, the cut-and-paste method was used to match these wormhole solutions onto an exterior flat Reissner--Nordstr{\"o}m spacetime.

    \item Turning to Figs.~\ref{fig4}, \ref{fig6}, and \ref{fig8}, the surface density ($\sigma_0$) turns out to be positive throughout all three thin--shell models, whereas the pressure $P_0$ does not retain a positive sign everywhere. For \textit{Model I}, $P_0$ is positive at $z=-0.5, ~0$, with mixed behavior elsewhere; for \textit{Model II}, only the $z=-0.5$ configurations yield a positive pressure, the remaining cases being negative; and for \textit{Model III}, the $z=-0.5$ curves are strictly positive, the $z=0.5$ curves are strictly negative, with the remaining case showing mixed behavior. Regarding the NEC, \textit{Model I} violates it solely for the $z=0.5$ configuration of the Joint dataset; \textit{Model II} violates NEC across all $z=0, ~0.5$ solutions; and in \textit{Model III}, the $z=-0.5$ solutions satisfy NEC while $z=0.5$ breaks it, with the remaining case again mixed. Because an NEC violation signals the appearance of exotic matter, the configurations in which such matter occurs can be readily identified. Since the remaining energy conditions are built directly upon the NEC, they naturally mirror its conclusions.

    \item Regarding stability, \cref{fig9} indicates that \textit{Model I} and \textit{Model II} are largely unstable, since the majority of curves fall below the $V''(y_0)=0$ line; in contrast, for \textit{Model III}, the thin--shells at $z=0.5$ come out stable while the remaining cases are unstable. In addition, stability improves as the junction radius $y_0$ increases, since most of the curves lying in the negative region increase monotonically with $y_0$.
\end{enumerate}

This concludes our study of GGDE-based wormholes within the framework of GR, in which three distinct models were investigated in detail across their various properties.

\section*{Acknowledgment}

SP and AK are thankful to IIEST, Shibpur, India, for providing Institute Fellowship (SRF).


\bibliographystyle{apsrev4-2}
\bibliography{2bibliography}


\end{document}